\documentclass[hyper,11pt,paper]{article}
\pdfoutput=1
\usepackage{jheppub}
\usepackage{tikz}
\usepackage{subcaption}

\newcommand{\p}{\partial}

\newcommand{\g}{\gamma}

\newcommand{\e}{\varepsilon}
\newcommand{\h}{\eta}
\renewcommand{\l}{\lambda}
\newcommand{\m}{\mu}
\newcommand{\n}{\nu}
\newcommand{\q}{\theta}
\newcommand{\f}{\phi}

\renewcommand{\r}{\rho}
\newcommand{\s}{\sigma}
\newcommand{\w}{\omega}

\newcommand{\N}{{\mathcal{N}}}
\newcommand{\vev}{\langle\textrm{tr} \,\Phi^2\rangle}
\newcommand{\beq}{\begin{equation}}
\newcommand{\eeq}{\end{equation}}
\newcommand{\bea}{\begin{eqnarray}}
\newcommand{\eea}{\end{eqnarray}}

\newcommand{\rhov}{\rho_{\textrm{v}}}

\newcommand{\cf}{\mathcal{F}}

\renewcommand{\le}{\left}
\newcommand{\ri}{\right}

\title{Holographic Wilson Lines as Screened Impurities}
\author[1]{Nick Evans,}
\author[1]{Andy O'Bannon,}
\author[1,2]{Ronnie Rodgers}

\affiliation[1]{STAG Research Centre, Physics and Astronomy, University of Southampton,\\Southampton, SO17 1BJ, United Kingdom}
\affiliation[2]{Institute for Theoretical Physics, Utrecht University, \\Princetonplein 5, 3584 CE Utrecht, the Netherlands}

\emailAdd{n.j.evans@soton.ac.uk}
\emailAdd{a.obannon@soton.ac.uk}
\emailAdd{r.j.rodgers@uu.nl}

\abstract{In Landau Fermi liquids, screened impurities support quasi-bound states, representing electrons bound to the impurity but making virtual excursions away. Signals of these quasi-bound states are electron-impurity scattering phase shifts and the corresponding resonances in cross sections. We consider large-$N$, strongly-coupled $(3+1)$-dimensional $\N=4$ supersymmetric $SU(N)$ Yang-Mills theory on the Coulomb branch, where an adjoint scalar has a non-zero expectation value that breaks $SU(N) \to SU(N-1) \times U(1)$. In the holographic dual we re-visit well-known solutions for a probe D3-brane that describe this theory with a symmetric-representation Wilson line ``impurity.'' We present evidence that the adjoint scalar screens the Wilson line, by showing that quasi-bound states form at the impurity, producing $U(1)$-impurity scattering phase shifts and corresponding resonances in cross sections. The quasi-bound states appear holographically as quasi-normal modes of probe D3-brane fields, even in the absence of a black hole horizon, via a mechanism that we argue is generic to screened defects in holography. We also argue that well-known generalisations of these probe D3-brane solutions can describe lattices of screened Wilson/'t Hooft line impurities.}

\begin{document}
	
\maketitle

\section{Introduction}
\label{sec:intro}

What effect does a quantum impurity, or a dilute concentration of quantum impurities, have on a Landau Fermi Liquid (LFL)? This is considered a ``solved problem'' thanks to a suite of complementary techniques, including the renormalization group (RG), Bethe ansatz, large-$N$ limits, Conformal Field Theory (CFT), and more. Typically, at sufficiently low concentration and low temperature $T$, each impurity's electric charge and/or magnetic moment will be screened by the conduction electrons, often leading to dramatic changes in the LFL's thermodynamic and transport quantities. For reviews, see for examples refs.~\cite{Hewson:1993,doi:10.1080/000187398243500,Phillips2012,Coleman2015,2016CRPhy..17..276V}.

Quite generally, quasi-bound states also form at the impurity: when the interaction between the impurity and the electrons (more precisely, LFL quasi-particles) is non-zero, the impurity spectral function develops a Lorentzian resonance whose residue and width are fixed by the coupling constant and the electronic density of states at the impurity's energy level~\cite{Hewson:1993,doi:10.1080/000187398243500,Phillips2012,Coleman2015,2016CRPhy..17..276V}. Physically, the resonance represents electrons bound to the impurity, or rather quasi-bound, since they can escape into the bulk. Indeed, the resonance's width arises from virtual excursions of electrons away from the impurity, into the LFL, and back.

However, what if the LFL electrons are replaced by strongly-interacting degrees of freedom? Do quasi-bound states form, and if so, what are their properties? Despite considerable progress using the techniques mentioned above, in general these problems remain unsolved.

In this paper we address these problems using the Anti-de Sitter (AdS)/CFT correspondence, also known as holography. Specifically, we consider $(9+1)$-dimensional type IIB supergravity (SUGRA) in $AdS_5 \times S^5$, where $AdS_5$ is $(4+1)$-dimensional AdS space and $S^5$ is a five-sphere. The holographic dual is $(3+1)$-dimensional $\N=4$ supersymmetric (SUSY) $SU(N)$ Yang-Mills (SYM) theory in the large-$N$ limit with large 't Hooft coupling~\cite{Maldacena:1997re,Gubser:1998bc,Witten:1998qj,Aharony:1999ti}.

We focus on well-known solutions for a probe D3-brane in $AdS_5 \times S^5$ that are holographically dual to 1/2-BPS Wilson lines in symmetric representations of $SU(N)$ on the Coulomb branch of $\N=4$ SYM~\cite{Gauntlett:1999xz,deMelloKoch:1999ui,Ghoroku:1999bc,Drukker:2005kx,Gomis:2006sb,Gomis:2006im,Schwarz:2014rxa}. In CFT terms, these solutions describe Wilson lines in states with a non-zero vacuum expectation value (VEV) of $\textrm{tr} \,\Phi^2$, with $\Phi$ an adjoint scalar of $\N=4$ SYM. On this Coulomb branch $SU(N)$ breaks to $SU(N-1) \times U(1)$, so the massless sector is $SU(N-1)$ $\N=4$ SYM plus $U(1)$ $\N=4$ SYM, and the lightest massive states are BPS multiplets bi-fundamental under $SU(N-1)\times U(1)$, which include the W-bosons.

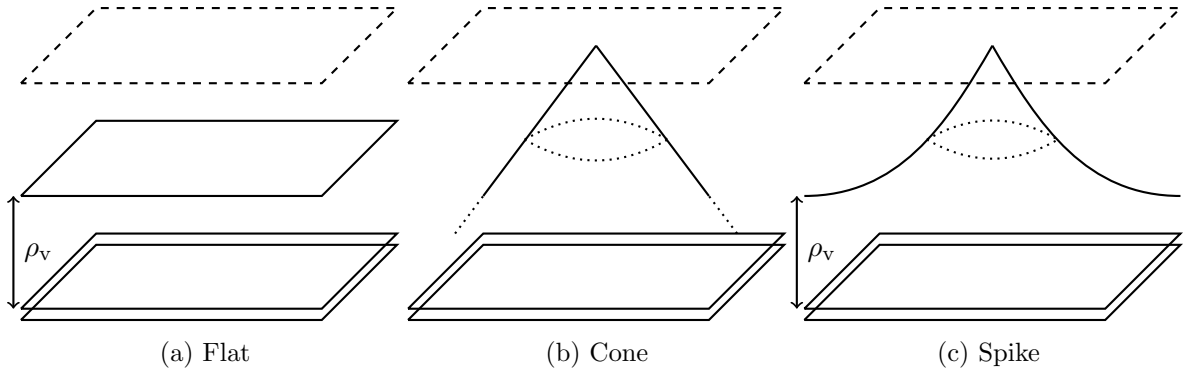
\begin{figure}
   \begin{subfigure}[t]{0.33\textwidth}
		\begin{tikzpicture}
		\draw[thick,dashed] (0,0)--(1,1)--(5,1)--(4,0)--(0,0);
		\draw[thick] (0,-1.5)--(1,-0.5)--(5,-0.5)--(4,-1.5)--(0,-1.5);
		\draw[thick] (0,-3)--(1,-2)--(5,-2)--(4,-3)--(0,-3);
		\draw[thick] (0,-3.15)--(1,-2.15)--(5,-2.15)--(4,-3.15)--(0,-3.15);
		\draw[thick, <->] (-0.1,-3)--(-0.1,-1.5) node[midway,right]{\(\rho_{\textrm{v}}\)};
		\end{tikzpicture}
		\caption{\label{flat}Flat}
	\end{subfigure}
	\begin{subfigure}[t]{0.325\textwidth}
		\begin{tikzpicture}
		\draw[thick,dashed] (0,0)--(1,1)--(5,1)--(4,0)--(0,0);
		\draw[thick] (1,-1.5)--(2.5,0.5);
		\draw[thick,dotted] (0.625,-2)--(1,-1.5);
		\draw[thick] (4,-1.5)--(2.5,0.5);
		\draw[thick,dotted] (4.375,-2)--(4,-1.5);
		\draw[thick] (0,-3)--(1,-2)--(5,-2)--(4,-3)--(0,-3);
		\draw[thick] (0,-3.15)--(1,-2.15)--(5,-2.15)--(4,-3.15)--(0,-3.15);
		\draw[thick,dotted] (1.55,-0.75) to[out=-30,in=210] (3.45,-0.75);
		\draw[thick,dotted] (1.55,-0.75) to[out=30,in=150] (3.45,-0.75);
		\end{tikzpicture}
		\caption{\label{cone}Cone}
	\end{subfigure}
		\begin{subfigure}[t]{0.33\textwidth}
		\begin{tikzpicture}
		\draw[thick,dashed] (0,0)--(1,1)--(5,1)--(4,0)--(0,0);
		\draw[thick] (0,-1.5) to[out=0,in=240] (2.5,0.5);
		\draw[thick] (5,-1.5) to[out=180,in=300] (2.5,0.5);
		\draw[thick] (0,-3)--(1,-2)--(5,-2)--(4,-3)--(0,-3);
		\draw[thick] (0,-3.15)--(1,-2.15)--(5,-2.15)--(4,-3.15)--(0,-3.15);
		\draw[thick,dotted] (1.63,-0.75) to[out=-30,in=210] (3.37,-0.75);
		\draw[thick,dotted] (1.63,-0.75) to[out=30,in=150] (3.37,-0.75);
		\draw[thick, <->] (-0.1,-3)--(-0.1,-1.5) node[midway,right]{\(\rho_{\textrm{v}}\)};
		\end{tikzpicture}
		\caption{\label{spike}Spike}
	\end{subfigure}
	\caption{\label{fig:branes}Cartoons of the probe D3-brane solutions we study in this paper. The parallelograms at the bottom represent the stack of D3-branes generating the background $AdS_5 \times S^5$, and the dashed parallelogram at the top represents the $AdS_5$ boundary. The shapes in between represent the probe D3-brane solutions describing (a) the $\N=4$ Coulomb branch, where the D3-brane's position $\rhov$ maps to an adjoint scalar's VEV, (b) a symmetric-representation 1/2-BPS Wilson line, and (c) such a Wilson line screened by the Coulomb branch's scalar VEV. We call these the ``flat,'' ``cone,'' and ``spike'' D3-brane solutions, respectively.}
\end{figure}

Figure~\ref{fig:branes} illustrates the probe D3-brane solutions we consider. In fig.~\ref{fig:branes} the vertical axis is the holographic direction $\rho$, with the $AdS_5$ boundary at $\rho \to \infty$ (top) and Poincar\'e horizon at $\rho=0$ (bottom). As an intuitive guide, at $\rho=0$ we depict the initial stack of coincident D3-branes that produce $AdS_5 \times S^5$, although these are not actually present: this SUGRA solution has $N$ units of five-form flux on the $S^5$ but no explicit D3-brane sources, i.e. the D3-branes ``dissolve'' into five-form flux. In fig.~\ref{fig:branes} we suppress the $S^5$. All the D3-brane solutions we consider sit at a point on the $S^5$, thus breaking its $SO(6)$ isometry to $SO(5)$.

Fig.~\ref{flat} depicts the probe D3-brane solution describing the Coulomb branch, with no Wilson line. Intuitively, we pull a single D3-brane from the stack of $N$ coincident D3-branes out to a constant value $\rhov$ (with $\textrm{v}$ for VEV) of the holographic coordinate, which maps to the symmetry-breaking scale $\vev \neq 0$. This solution clearly breaks $SU(N) \to SU(N-1) \times U(1)$, with the $U(1)$ being the probe D3-brane's worldvolume gauge invariance. Fluctuations of this D3-brane's worldvolume fields represent fluctuations of the $U(1)$ $\N=4$ SYM fields at scales below the symmetry breaking scale $\vev$. This D3-brane also breaks the $AdS_5$ isometries to those of a Poincar\'e slice at fixed $\rho$, dual to $\vev \neq 0$ breaking the conformal group to the Poincar\'e group. Due to the Poincar\'e symmetry we call this a ``flat'' D3-brane. 

Fig.~\ref{cone} depicts the probe D3-brane solution describing a 1/2-BPS Wilson line in a symmetric representation of $SU(N)$, with $\vev =0$. In the CFT, the Wilson line is a static, point-like conformal defect that breaks the conformal group $SO(4,2)$ to $SO(1,2) \times SO(3)$, where $SO(1,2)$ are conformal transformations leaving the Wilson line invariant (time translations, dilatations, and special conformal transformations involving inversions through the Wilson line's location), and $SO(3)$ are rotations around the Wilson line. Correspondingly, the probe D3-brane's worldvolume is $AdS_2 \times S^2$, whose isometries are $SO(1,2)$ and $SO(3)$, respectively. The $AdS_2$'s $(0+1)$-dimensional boundary is at the $AdS_5$ boundary, at the Wilson line's location. Graphically the D3-brane looks like a ``cone'' with apex at the $AdS_5$ boundary whose opening angle determines the dimension of the Wilson line's $SU(N)$ representation.

Fig.~\ref{spike} depicts the probe D3-brane ``spike'' solution that interpolates between the flat and cone D3-branes of figs.~\ref{flat} and~\ref{cone}. At spatial infinity in any field theory direction these solutions approach the flat D3-brane sitting at a fixed value of $\rho$, representing a point on the Coulomb branch. However,  a spike emerges from the D3-brane and reaches the $AdS_5$ boundary at a point in the field theory directions. In the near-boundary region the worldvolume metric approaches $AdS_2 \times S^2$ and the D3-brane resembles the cone solution.

These spike solutions are well known, having been re-discovered many times in the last 20 years, in various contexts~\cite{Gauntlett:1999xz,deMelloKoch:1999ui,Ghoroku:1999bc,Drukker:2005kx,Fiol:2008gt,Schwarz:2014rxa,Schwarz:2014zsa}. However, to our knowledge no dual field theory interpretation has been proposed for them. We propose interpreting them as Wilson line ``impurities'' screened by the Coulomb branch VEV. To see why, simply recall that in the holographic dictionary $\rho$ corresponds to the energy scale of the dual CFT, with the ultra-violet (UV) near the $AdS_5$ boundary and infra-red (IR) near the Poincar\'e horizon. In these solutions, clearly the impurity is present in the UV but absent in the IR, because $\vev$ has adjusted itself in space to screen the impurity, in a way that preserves spherical symmetry around the impurity. In other words, these solutions represent a spherically-symmetric screening cloud made of adjoint scalar VEV. The $SU(N)$ adjoint is essentially the combination of the fundamental and anti-fundamental representations, so intuitively we can think of the screening cloud as a collection of adjoint scalar ``dipoles'' polarized by the impurity.

In fact, another solution exists with the same boundary conditions as the spike, both at the $AdS_5$ boundary and at spatial infinity in the field theory, namely a superposition of the flat and cone D3-branes. (This solution may only exist in the probe limit, where these two D3-branes do not interact.) Per the usual holographic dictionary, the solution with smaller on-shell action will be preferred. However, both of these solutions are SUSY and so have vanishing action, hence neither is preferred over the other. The corresponding field theory statement is that the scalar VEV can screen the impurity without changing the energy from zero, i.e. no loss or gain of energy occurs. This is clearly a special feature of SUSY, since in most impurity systems without SUSY, typically at $T=0$ screening is energetically favoured.

To support our interpretation of the spike solution as a screened impurity, we will present ``smoking gun'' evidence, namely quasi-bound states. In SUGRA terms, we will compute the spectrum of linearised fluctuations of D3-brane worldvolume fields about the spike solution and find quasi-normal modes (QNMs) holographically dual to massless $U(1)$ sector degrees of freedom quasi-bound to the Wilson line impurity.

Crucially, since $T = 0$ these QNMs arise not from the physics of horizons, but via a mechanism that to our knowledge is novel. We need two boundary conditions for our linearised worldvolume fluctuations. Near the spike's tip, where the worldvolume is asymptotically $AdS_2 \times S_2$, we impose normalisability. This corresponds to injecting energy by fluctuating a source at the impurity. Infinitely far from the spike's tip the worldvolume is asymptotically flat, hence we find in- and out-going wave solutions. Here we require fluctuations to be purely out-going, so that the energy we injected flows out of the system at infinity, but no energy flows in~\cite{Berti:2009kk}. Although our linearised fluctuation operator is Hermitian, the latter boundary condition is non-Hermitian, thus giving us QNMs instead of normal modes. In short, our injected energy leaks out at asymptotic infinity in the field theory directions, in contrast to black hole QNMs, where energy is injected from infinity and absorbed by the horizon.

Such a mechanism is clearly quite general, and will occur in any holographic system where a continuum of bulk modes can scatter off a defect whose geometry is not simply $AdS$. In field theory terms, such a mechanism will occur whenever a continuum of modes scatters of a non-conformal defect, where the breaking of defect conformal symmetry is necessary because a scale is needed to set the spacing between QNMs. We expect the same mechanism to occur for example with any probe defect whose worldvolume is not simply AdS but is asymptotically flat, like the spike D3-brane, or whose worldvolume is not AdS and not necessarily asymptotically flat, but back-reacts, so that SUGRA modes can scatter off it. As so often before, holography thus provides a very intuitive geometric picture of a generic phenomenon: non-conformal defects will generically produce quasi-bound states of bulk modes.

We find several other generic features as well. We put the linearized fluctuation equations into the form of Schr\"{o}dinger equations with potentials determined by the D3-brane's worldvolume geometry. The QNMs then correspond to meta-stable states, many of which are quasi-bound in finite wells of these potentials. We then perform a standard quantum mechanical scattering analysis: we scatter massless $U(1)$ sector degrees of freedom off the impurity and determine the resulting s- and p-wave phase shifts and associated cross sections. The meta-stable states produce rapidly changing phase shifts and peaks in the cross sections whose locations and widths are determined by the QNMs. In other words, we inject energy from infinity in field theory directions, and rather than being absorbed by a horizon, the energy is reflected off the impurity, with signatures of the QNMs. The cross section peaks are in fact Fano resonances, that is, asymmetric line-shapes that occur whenever a continuum scatters off a localized resonance~\cite{RevModPhys.82.2257}. Our Fano resonances are perfect examples of the mechanism discovered in refs.~\cite{Erdmenger:2013dpa,Erdmenger:2016vud,Erdmenger:2016jjg}, because they arise from the breaking of $(0+1)$-dimensional conformal symmetry, in our case at the Wilson line impurity.

This paper is organized as follows. In section~\ref{sec:spike} we review the D3-brane spike solution. In sec.~\ref{sec:fluc} we determine the equations of motion of worldvolume fluctuations about the spike. In sec.~\ref{sec:qnms} we compute the QNM spectrum. In sec.~\ref{sec:scattering} we compute the phase shifts and cross sections.  We conclude in sec.~\ref{sec:outlook} with a summary, and suggestions for future research.

\section{Probe D3-brane Spike Solutions}
\label{sec:spike}

We will be interested in probe D3-branes in the $AdS_5 \times S^5$ solution of type IIB SUGRA. We write the metric $G_{MN}$ with $M,N=0,1,\ldots,9$ and four-form $C_{(4)}$ as
\begin{subequations}
\begin{align}
	d s^2 &= G_{MN} dx^Mdx^N = \frac{\rho^2}{L^2} \, \h_{\m\n} d x^\m d x^\n + \frac{L^2}{\rho^2} \left( d \rho^2 + \rho^2 ds_{S^5}^2 \right),
	\\
	C_{(4)} &= \frac{\rho^4}{L^4} \, d t \wedge d x^1 \wedge d x^2 \wedge d x^3,
\end{align}
\end{subequations}
where $\rho$ is the holographic coordinate, with $\rho = 0$ the Poincar\'{e} horizon and $\rho \to \infty$ the $AdS_5$ boundary, $L$ is the $AdS_5$ radius of curvature, $x^{\mu}=(t,x^1,x^2,x^3)$ are the CFT coordinates, and $ds_{S^5}^2$ is the metric of a round unit-radius $S^5$.

We will need only the bosonic part of the D3-brane action, $S_{D3}$, whose bulk part (ignoring boundary terms) is a sum of Dirac-Born-Infeld (DBI) and Wess-Zumino (WZ) terms,
\begin{subequations}
\label{eq:d3action}
\beq
S_{D3} = S_\mathrm{DBI} + S_\mathrm{WZ},
\eeq
\beq
S_{\textrm{DBI}} = - T_{D3} \int d^4 \xi \sqrt{-\det(g+ F)}, \qquad S_{\textrm{WZ}} =  T_{D3} \int P[C_{(4)}],
\eeq
\end{subequations}
where $T_{D3}=(2\pi)^{-3} g_s^{-1} \alpha'^{-2}$ is the D3-brane tension, with string coupling $g_s$ and string length squared $\alpha'$, $\xi^a$ with $a = 0,1,2,3$ are the worldvolume coordinates, $g_{ab} = \p_a X^M \p_b X^N G_{MN}$ and $P[C_{(4)}]$ are the pullbacks to the D3-brane of the bulk metric and four-form, respectively, with D3-brane worldvolume scalars $X^M$, and $F_{bc}$ is the field strength of the $U(1)$ worldvolume gauge field $A_b$, which we have made dimensionless by absorbing a factor of $(2\pi\alpha')$.

To obtain the flat, cone, and spike solutions we choose a gauge $\xi^{\mu} = x^{\mu}$ and look for static solutions in which $\rho$ and $A_t$ are the only non-zero worldvolume fields, and furthermore are spherically symmetric and so depend only on the CFT radial coordinate $r \equiv \sqrt{(x^1)^2 + (x^2)^2 + (x^3)^2}$. For such an ansatz the worldvolume metric is,
\beq
\label{eq:wvmet}
g_{ab} \, d\xi^a d\xi^b = \frac{\rho^2}{L^2}\left[-dt^2+\left(1+\frac{L^4}{\rho^4}\,\rho'^2\right)dr^2+r^2\,ds^2_{S^2}\right],
\eeq
where $\rho' \equiv \partial_r \rho$ and $ds^2_{S^2}$ is the metric of a round, unit-radius $S^2$,
\beq
\label{eq:s2metric}
ds^2_{S^2} = d \q^2 + \sin^2\q \, d\f^2,
\eeq
where $\theta \in [0,\pi]$ and $\phi\in[0,2\pi]$. The only non-zero component of the field strength is a radial electric field, $F_{tr}$. We choose a $U(1)$ gauge in which $A_r=0$, so that $F_{tr}  = - \partial_r A_t \equiv -A_t'$. The D3-brane action then takes the form
\beq
\label{simpleaction}
S_{D3} = - 4\pi\,T_{D3} \int  \, dr \, r^2 \left [ {\rho^4 \over L^4}\sqrt{1 +  \left. {L^4 \over \rho^4} \left[ \rho^{'2} - (A_t')^2  \right]  \right. } - \frac{\rho^4}{L^4} \right].
\eeq
The equations of motion that follow from this action have the 1/2-BPS solution~\cite{Gauntlett:1999xz,deMelloKoch:1999ui,Ghoroku:1999bc,Drukker:2005kx,Fiol:2008gt,Schwarz:2014rxa,Schwarz:2014zsa}
\beq
\label{basicspike}
\rho = \rhov + \frac{L^2\,Q}{r},  \qquad A_t =  \rho,
\eeq
where $\rhov>0$ and $Q>0$ are integration constants, of dimensions $(\textrm{length})$ and $(\textrm{length})^0$, respectively. This solution is the global minimiser of the $S_{D3}$ in eq.~\eqref{simpleaction}, and indeed $S_{D3}$ vanishes when evaluated on this solution, as required by SUSY~\cite{deMelloKoch:1999ui} (the boundary terms in the action also vanish when evaluated on this solution). In this solution the D3-brane sits at a point on the $S^5$, thus breaking the corresponding $SO(6)$ isometry down to $SO(5)$.

Our goal is to compute the spectrum of linearised fluctuations of D3-brane (bosonic) worldvolume fields around the solution in eq.~\eqref{basicspike}. Results for this spectrum are already known in two limits, namely the flat and the cone D3-brane solutions, obtained by setting one or the other integration constant to zero in eq.~\eqref{basicspike}.

The flat D3-brane has $\rhov\neq0$ and $Q=0$. In that case the worldvolume metric in eq.~\eqref{eq:wvmet} is simply a $(3+1)$-dimensional Minkowski metric times a constant overall factor $\rhov^2/L^2$, and $A_t = \rhov$ is constant, leading to a vanishing field strength, $F_{tr}=0$. The linearised fluctuations of all worldvolume fields are then simply plane waves with fixed momenta.

As mentioned in sec.~\ref{sec:intro}, the flat D3-brane solution represents a non-zero adjoint scalar VEV, $\vev\propto \rhov^2/L^4 \neq0$, which breaks $SU(N) \to SU(N-1)\times U(1)$, producing $SU(N-1)$ and $U(1)$ $\N=4$ SYM multiplets coupled via a W-boson multiplet with mass $\rhov/(2\pi\alpha')$. Crucially, for the flat D3-brane's worldvolume fields the ``usual'' holographic dictionary does not apply. ``Usual'' means: solve for fields as functions of $\rho$, series expand about the $AdS_5$ boundary (i.e. in powers of $1/\rho$), and then identify the leading, non-normalizable terms as sources and the sub-leading, normalizable terms as VEVs. The flat D3-brane sits at a fixed value of $\rho$, so such a procedure is clearly inapplicable. Instead, we directly identify the flat D3-branes' worldvolume fields as those of the $U(1)$ $\N=4$ SYM multiplet, and the D3-brane action as their effective action obtained by integrating out the W-boson multiplet~\cite{Schwarz:2013wra,Schwarz:2014rxa,Schwarz:2014zsa}. The linearised fluctuation of the D3-brane fields are thus not dual to poles in retarded Green's functions, rather they are identically the fluctuations of the $U(1)$ $\N=4$ SYM fields.

The cone D3-brane has $\rhov=0$ and $Q \neq 0$. In that case, after re-scaling $t$ as
\beq
t \quad \rightarrow \quad  \hat{t} \equiv t\,\left(1+Q^{-2}\right)^{-1/2},
\eeq
the D3-brane worldvolme metric in eq.~\eqref{eq:wvmet} becomes that of $AdS_2 \times S^2$,
\beq
\label{eq:ads2s2}
g_{ab} \, d\xi^a d\xi^b = L^2 \left(1+ Q^{2}\right)\left(\frac{-d\hat{t}^2+dr^2}{r^2}\right)+Q^2 L^2\,ds^2_{S^2},
\eeq
where the $AdS_2$ radius is $L\sqrt{1+Q^{2}}$ and the $S^2$ radius is $Q L$. The cone D3-brane intersects the $AdS_5$ boundary at the point $r=0$, which on the worldvolume is the $AdS_2$ boundary. This solution has non-vanishing field strength $F_{tr} = Q L^2/r^2$, indicating that the D3-brane carries non-vanishing string charge $4 N Q/\sqrt{\lambda}$. Heuristically, we can imagine that these solutions represent strings ending on the $AdS_5$ boundary that have ``puffed up'' into a D3-brane via a Myers effect~\cite{Myers:1999ps}.  A single string with $AdS_2$ worldsheet is dual to a 1/2-BPS Wilson line in the fundamental representation of $SU(N)$~\cite{Rey:1998ik,Maldacena:1998im}. Correspondingly, the cone D3-brane is dual to a 1/2-BPS Wilson line at $r=0$ in a symmetric representation of $SU(N)$ whose Young tableau has a number of boxes $4 N Q/\sqrt{\lambda}$~\cite{Rey:1998ik,Drukker:2005kx,Gomis:2006sb,Gomis:2006im}.

The $SO(1,2)$ isometry of the worldvolume $AdS_2$ factor indicates that the dual 1/2-BPS Wilson line preserves the $SO(1,2)$ subgroup of the $SO(4,2)$ conformal group that leaves the Wilson line's position invariant. In other words, the 1/2-BPS Wilson line is a conformal defect. As a result, two-point functions of operators localized on the Wilson line are completely determined by the operators' dimensions and charges under the $SO(3)$ rotational symmetry and $SO(5)$ R-symmetry. Via holography, the corresponding statement is that the linearised fluctuations of worldvolume fields are completely determined by their masses and charges alone, as shown in ref.~\cite{Faraggi:2011bb}.

The spike D3-brane is the solution with both $\rhov\neq0$ and $Q\neq0$. As mentioned in sec.~\ref{sec:intro}, we will interpret this solution as a 1/2-BPS symmetric-representation Wilson line ``impurity'' screened by the Coulomb branch VEV in a spherically-symmetric fashion~\cite{Gauntlett:1999xz,deMelloKoch:1999ui,Ghoroku:1999bc,Drukker:2005kx,Gomis:2006sb,Gomis:2006im,Schwarz:2014rxa}. Indeed, the worldvolume metric of this solution interpolates from the flat D3-brane's Minkowski metric far from the spike, $r \to \infty$, to the cone D3-brane's $AdS_2 \times S^2$ metric near the spike, $r \to 0$. The worldvolume field strength is the same as the cone D3-brane, $F_{tr} = Q L^2/r^2$.

In subsequent sections we will study the spectrum of linearised fluctuations of (bosonic) worldvolume fields on the spike D3-brane. The results summarised above will appear as limits: far from the spike, $r \to \infty$, the fluctuations will reduce to the flat D3-brane's plane waves, while near the spike, $r \to 0$, we will find the cone D3-brane spectrum of ref.~\cite{Faraggi:2011bb}, which is fully determined by the fluctuations' masses and charges. To obtain QNMs, in the $r \to \infty$ region we will impose that the fluctuations are purely out-going, and in the $r \to 0$ region we will impose that the fluctuations are normalisable in $AdS_2$. In physical terms, we will inject energy through the impurity, which then ``leaks out'' to spatial infinity in CFT directions.

The spike D3-brane reaches the $AdS_5$ boundary, hence the usual holographic dictionary applies: in principle, the spike D3-brane solution actually represents operators localized on the Wilson line that have acquired $SO(3)$-symmetric VEVs that ``mimic'' the effects of $\vev \neq 0$. However we will not pursue such an interpretation. Instead, the results of the following sections will suffice to justify the more intuitive interpretation that the Coulomb branch VEV has screened the Wilson line impurity. These two interpretations must be equivalent, but we will leave the task of proving so for future research.

Although in this work we restrict to solutions with \(\rho_\mathrm{v} > 0\), the solution eq.~\eqref{basicspike} is also valid if \(\rho_\mathrm{v} < 0\)~\cite{Kumar:2016jxy}. When \(\rho_\mathrm{v}\) is negative the D3-brane does not flatten out, but instead reaches all the way to \(\rho = 0\). At small \(\rho\) the worldvolume of the brane is a warped product of \(AdS_2\) and \(S^2\), with the \(S^2\) shrinking to zero size as \(\rho \to 0\). The D3-brane thus becomes effectively two-dimensional in the deep IR, and resembles a marginally bound state of \(4 N Q/\sqrt{\l}\) fundamental strings. This solution is holographically dual to a symmetric representation Wilson line screened by a non-zero VEV of an impurity-localised operator, such that in the IR the impurity looks like \(4 N Q/\sqrt{\l}\) coincident fundamental representation Wilson lines. Similar to the spike solution, SUSY implies that this form of screening may occur with no loss or gain of energy.

A more general 1/2-BPS solution is known, namely a multi-centre solution with both electric and magnetic fields~\cite{Gauntlett:1999xz,deMelloKoch:1999ui}. This more general solution is static but not spherically symmetric, depending in general on all of the CFT spatial coodinates $x^i$ with $i=1,2,3$. In this solution $\rho$ is determined by the Laplace equation in three-dimensional Euclidean space,
\beq
\label{eq:laplace}
\delta^{ij} \p_i \p_j \rho = 0,
\eeq
while the electric fields $F_{ti}$ and magnetic fields $F_{jk}$ are completely determined by $\rho$ via vector-scalar duality conditions,
\beq
\label{eq:vector_scalar}
F_{ti} = \cos\chi \, \p_i \rho,
\qquad \frac{1}{2} \e_{ljk} F^{jk} = \sin\chi \, \p_i \rho,
\eeq
where $\e_{ijk}$ is the Levi-Civita symbol (with $\e_{123}=1$), and $\chi \in [0,2\pi)$ is a free parameter. Eq.~\eqref{eq:laplace} shows that finding multi-centre solutions reduces to a problem in three-dimensional electrostatics. In particular, any linear combination of solutions to eq.~\eqref{eq:laplace} is again a solution, hence we can construct multi-centre solutions simply by taking linear combinations of the single spike solution in eq.~\eqref{basicspike}, placing the spikes anywhere we like on the D3-brane worldvolume. The presence of both electric and magnetic fields indicates that these spikes will generically be dual to mixed Wilson/'t Hooft lines. Using such multi-centre solutions we can thus construct lattices of impurities, including lattices with broken discrete symmetries, for example lattices of magnetic impurities that break time reversal. A natural question then is whether such lattices produce non-trivial band structure, and in particular whether they support (topologically-protected) massless modes. We return to this question in sec.~\ref{sec:outlook}.

\section{Linearised Fluctuations}
\label{sec:fluc}

We now consider linearised fluctuations of D3-branes worldvolume fields around the spike solution eq.~\eqref{basicspike}. The spike solution preserves SUSY, and more specifically is 1/2-BPS, hence the linearised fluctuations will form SUSY multiplets. We will focus on the bosonic fluctuations, which include the six scalars describing the D3-brane's position in transverse directions and the four components of the $U(1)$ worldvolume gauge field.

In fact, we will focus only on a subset of these bosonic fluctuations. The spike solution preserves $SO(3)$ rotations around the impurity, as well as an $SO(5)$ R-symmetry. At the linearised level, fluctuations in different $SO(3) \times SO(5)$ representations will decouple. We will focus only on $SO(5)$ singlets, which include the scalar fluctuation describing the D3-brane's position in the holographic coordinate $\rho$ and the four $U(1)$ gauge field components.

We thus introduce the small fluctuations about the spike solution eq.~\eqref{basicspike},
\begin{equation}
	\r(r) \to \r(r) + \delta \rho(t,r,\q,\f),
	\quad
	A_b(r) \to A_b(r) + \delta A_b(t,r,\q,\f),
\end{equation}
with $\delta \rho \ll \rho$ and $\delta A_b \ll A_b$. The fluctuations $\delta \rho$, $\delta A_t$, and $\delta A_r$ are $SO(3)$ scalars, which following ref.~\cite{Faraggi:2011bb} we expand in $S^2$ scalar harmonics $Y^{lm}(\q,\f)$ with coefficients that depend only on $t$ and $r$,
\begin{subequations}
\label{eq:vector_harmonic_expansion1}
\begin{align}
	\delta \rho(t,r,\q,\f) &= \sum_{l=0}^\infty \sum_{m=-l}^l \rho_{lm}(t,r) Y^{lm}(\q,\f),
	\\
	\delta A_t(t,r,\q,\f) &=  \sum_{l=0}^\infty \sum_{m=-l}^l A_t^{lm}(t,r) Y^{lm}(\q,\f),
	\\
	\delta A_r(t,r,\q,\f) &=  \sum_{l=0}^\infty \sum_{m=-l}^l A_r^{lm}(t,r) Y^{lm}(\q,\f).
\end{align}
\end{subequations}
The fluctuations of the $S^2$ components of the $U(1)$ gauge field, $\delta A_{\beta}$ with $\beta = (\theta,\phi)$ the $S^2$ coordinates in eq.~\eqref{eq:s2metric}, are an $SO(3)$ vector, which we expand in $S^2$ vector harmonics. The complete set of vector-valued functions on $S^2$ includes derivatives of scalar harmonics, $\partial_{\beta} Y^{lm}$, and derivatives of scalar harmonics combined with a Levi-Civita symbol, $\e^{\beta\gamma}$ (with $\e^{\q\f} = 1$),
\beq
\hat{Y}^{lm}_{\beta}(\q,\f) \equiv \frac{1}{\sqrt{l(l+1)}} \,{\e_{\beta}}^{\gamma} \, \p_{\gamma} Y^{lm}(\q,\f),
\eeq
where we raise and lower indices with the $S^2$ metric. However, as explained in ref.~\cite{Faraggi:2011bb} we can use a partial gauge fixing to remove the former, hence we expand $\delta A_{\beta}$ only in the latter, again with coefficients that depend only on $t$ and $r$,
\beq
\label{eq:vector_harmonic_expansion2}
\delta A_{\beta}(t,r,\q,\f) = \sum_{l=1}^\infty \sum_{m=-l}^l B_{lm}(t,r) \hat Y_{\beta}^{lm}(\q,\f).
\eeq

We next insert eqs.~\eqref{eq:vector_harmonic_expansion1} and~\eqref{eq:vector_harmonic_expansion2} into $S_{D3}$ in eq.~\eqref{eq:d3action} and expand, obtaining
\beq
S_{D3} = S_{D3}^{(0)} + S_{D3}^{(1)} + S_{D3}^{(2)} + \ldots
\eeq
where the superscripts denote powers of fluctuations, so $S_{D3}^{(0)}$ is simply eq.~\eqref{simpleaction} evaluated on the solution in eq.~\eqref{basicspike}, which vanishes due to SUSY as mentioned above. The equations of motion require $S_{D3}^{(1)}=0$. The linearised equations of motion for the fluctuations come from $S_{D3}^{(2)}$. To write $S_{D3}^{(2)}$ succinctly, we define 
\beq
H \equiv L^4/\rho^4, \qquad \mathcal{F}_{tr}^{lm} \equiv \p_t A_r^{lm} - \p_r A_t^{lm}, \qquad (\rho_{lm})^2 \equiv \rho_{lm} \rho_{l,-m},
\eeq
and similarly for $(\mathcal{F}_{tr}^{lm})^2$, $\left(A_t^{lm}\right)^2$, etc. We find
\begin{align}
\label{eq:kk_reduced_action}
S_{D3}^{(2)} 
&= -T_3 \sum_{l,m} (-1)^m\int d t d r \, r^2\biggl\{
- \frac{1}{2} (\p_t \rho_{lm})^{2}
+ \frac{1 - H \rho'^2}{2} \left[(\p_r \rho_{lm})^2  +  \frac{l(l+1)}{r^2} (\rho_{lm})^2\right]
\nonumber \\ 
&\phantom{-T_3 \sum_{l,m} (-1)^m\biggl\{}
- \frac{1 + H \rho'^2}{2} 
(\mathcal{F}_{tr}^{lm})^2
- \frac{1 + H \rho'^2}{2r^2} l(l+1) (A_t^{lm})^2+ \frac{1}{2 r^2} l (l+1) (A_r^{lm})^2
\nonumber \\
&\phantom{-T_3 \sum_{l,m} (-1)^m\biggl\{}
+ H \rho'^2 \mathcal{F}_{tr}^{lm} \p_r \rho_{l,-m} - \frac{H \rho'^2}{r^2} l(l+1) A_t^{lm} \rho_{l,-m} 
\nonumber\\
&\phantom{-T_3 \sum_{l,m} (-1)^m\biggl\{}
- \frac{1 + H \rho'^2}{2r^2} (\p_t B_{lm})^2 + \frac{1}{2r^2} (\p_r B_{lm})^2
+ \frac{1}{2 r^4} l(l+1) (B_{lm})^2 + \ldots \biggr\},
\end{align}
where the $\ldots$ represents the terms involving the $SO(5)$ vector fluctuations, which as mentioned above do not couple to the fluctuations we are considering. At this linearised level, only fluctuations with the same values of $l$ and $|m|$ couple to one another. In particular, the scalar harmonic coefficients $\rho_{lm}$, $A_t^{lm}$ and $A_r^{lm}$ couple to one another, via the terms in the third line of eq.~\eqref{eq:kk_reduced_action}, but decouple from the vector harmonic coefficients $B_{lm}$, as expected.

Since the equations of motion for the fluctuations do not depend on $m$, starting now we will consider only fluctuations with $m=0$ without loss of generality. We define the notation $\rho_{lm=0} \equiv \rho_l$ and similarly for the other fluctuations. The equations of motion for $\rho_l$, $A_t^l$, and $A_r^l$ derived from $S_{D3}^{(2)}$ in eq.~\eqref{eq:kk_reduced_action} are then, respectively,
\begin{subequations}
\label{eq:eoms1}
\beq
\label{eq:rhoeom}
\p_r \le[r^2 \le(1 -  H \r'^2 \ri) \p_r \rho_l \ri] - r^2 \p_t^2 \rho_l- l(l+1) \le(1 - H \r'^2 \ri) \rho_l + \p_r \le(r^2 H \r'^2 \cf_{tr}^l\ri) + l(l+1) H \r'^2 A_t^l = 0,	
\eeq
\beq
\label{eq:ateom}
\p_r \le[r^2 \le(1 + H \r'^2 \ri) \cf_{tr}^l  - r^2 H \r'^2 \p_r \rho_l \ri] + l(l+1) \le(1 + H \r'^2 \ri) A_t^l + l(l+1) H \r'^2 \rho_l = 0,
\eeq
\beq
\label{eq:areom}
\p_t \le[r^2 \le(1 + H \r'^2 \ri) \cf_{tr}^l  - r^2 H \r'^2 \p_r \rho_l \ri] + l(l+1) A_r^l = 0.
\eeq
\end{subequations}
Henceforth we will also focus only on a single Fourier mode in time of each fluctuation, meaning we will take $\rho_l(t,r) \equiv e^{-i \w t} \rho_l(r)$ with frequency $\omega$, and similarly for $A_t^l$ and $A_r^l$. (We use the same symbol for the fluctuation and its Fourier mode. The difference should always be clear from the context.) Using the definition $\cf_{tr}^l \equiv - i \omega A_r^l - \partial_r A_t^l$ we can solve eq.~\eqref{eq:areom} for $A_r^l$, and then plug the result into eqs.~\eqref{eq:rhoeom} and~\eqref{eq:ateom}, thus obtaining two coupled equations for $\rho_l$ and $A_t^l$ alone. Suitable linear combinations of those equations produce new equations that are almost (though not quite) symmetric under the exchange of $\rho_l$ and $A_t^l$,
\begin{subequations}
\label{eq:leoms}
\begin{align}
\label{eq:rholeom}
	\partial_r^2 \rho_l + \frac{\le[l(l+1) - \w^2 r^2\ri] \le[ r \p_r \le(H \r'^2 \ri) - 2\ri] + 2 \w^2 r^2 H \r'^2}{r \le[ \w^2 r^2 (1 + H \r'^2) - l(l+1) \ri]} \partial_r \rho_l
	\hspace{2.8cm} & \nonumber \\
	+ \frac{l(l+1) \p_r(H \r'^2 )}{\w^2 r^2 (1 + H \r'^2) - l(l+1)} \partial_rA_t^l
	+ \le[ \w^2 (1 + H \r'^2) - \frac{l(l+1)}{r^2} \ri] \rho_l &= 0,
	\\[1em]
	\label{eq:atleom}
	\partial_r^2A_t^l - \frac{\le[ l(l+1) - \w^2 r^2 \ri] \p_r \le( H \r'^2 \ri) +2 \w^2 r H \r'^2}{ \w^2 r^2 (1 + H \r'^2) - l(l+1)} \partial_r \rho_l
	\hspace{4cm} & \nonumber \\
	- \frac{l(l+1) \le[ r \p_r (H \r'^2) + 2 \ri]}{r \le[ \w^2 r^2 (1 + H \r'^2) - l(l+1) \ri]} \partial_rA_t^l + \le[ \w^2 (1 + H \r'^2) - \frac{l(l+1)}{r^2} \ri] A_t^l &= 0.
\end{align}
\end{subequations}
Solutions of eq.~\eqref{eq:leoms} have asymptotic expansions around the spike at $r=0$ with forms characteristic of asymptotically $AdS_2$ spacetime, as expected,
\begin{subequations}
\label{eq:eomsmallr}
\begin{align}
\label{eq:rhosmallr}
	\rho_l &= \frac{c^{\rho}_{-l-1}}{r^{l+1}} \, \le[1 + \mathcal{O}\left(r\right)\ri] + d^{\rho}_l \, r^l \le[1  + \mathcal{O}\left(r\right) \ri], \\
	\label{eq:atsmallr}
	A_t^l &=  \frac{c^t_{-l-1}}{r^{l+1}} \, \le[1 +  \mathcal{O}\left(r\right) \ri] + d^t_l \, r^l \le[1 +  \mathcal{O}\left(r^2\right) \ri],
\end{align}
\end{subequations}
where the complex-valued coefficients $c^{\rho}_{-l-1}$, $d^{\rho}_l$, $c^t_{-l-1}$, and $d^t_l$ are integration constants, meaning they are independent of $r$ and completely determine the coefficients of all other powers of $r$ in eq.~\eqref{eq:eomsmallr}. The asymptotic expansions of the solutions around spatial infinity $r\to \infty$ have forms characteristic of scalar harmonics in Minkowski spacetime, as expected,
\begin{subequations}
\label{eq:eomlarger}
\begin{align}
	\rho_l &=  \frac{f_l^{\rho}(r)}{r}\,e^{-i (\w r - l \pi /2)} + \frac{g_l^{\rho}(r)}{r}\,
		e^{i (\w r - l \pi /2)},
	 \\
	A_t^l &=
		f_l^t(r)\,e^{-i (\w r - l \pi /2)}
		+ g_l^t(r)\,e^{i (\w r - l \pi /2)},
\end{align}
\end{subequations}
where the complex-valued functions $f_l^{\rho}(r)$, $g_l^{\rho}(r)$, $f_l^t(r)$, and $g_l^t(r)$ are regular in $r$ as $r \to \infty$.

For the vector harmonics $B_l$, the equation of motion for a single Fourier mode is
\beq
\label{eq:abetaeom}
\p_r^2 B_l + \left[ \omega^2 \le(1 + H \r'^2 \ri) - \frac{l(l+1)}{r^2} \right] B_l = 0.
\eeq
A solution of eq.~\eqref{eq:abetaeom} has an asymptotic expansion around $r=0$ with a form characteristic of asymptotically $AdS_2$ spacetime, as expected,
\beq
\label{eq:abetasmallr}
B_l = \frac{c^B_{-l}}{r^l} \,\le[1 +  \mathcal{O}\left(r^2\right) \ri] + d^B_{l+1} \, r^{l+1} \, \le[1 +  \mathcal{O}\left(r^2\right) \ri],
\eeq
with integration constants $c^B_{-l}$ and $d^B_{l+1}$. The asymptotic expansion of a solution around $r \to \infty$ has the form characteristic of a vector harmonic in Minkowski spacetime, as expected,
\beq
\label{eq:abetalarger}
B_l = f_l^B(r) \,e^{-i (\w r - l \pi/2)} + g_l^B(r)\,e^{i (\w r - l \pi/2)} ,
\eeq
where the two functions $f_l^B(r)$ and $g_l^B(r)$ are regular in $r$ as $r \to \infty$.

Perfectly spherical fluctuations are special: setting $l=0$ eliminates $A_t^0$ from eq.~\eqref{eq:rholeom}, which thus becomes an equation for $\rho_0$ alone,
\beq
\label{eq:rho0eom}
\partial_r^2 \rho_0 + \left[\partial_r\ln\left(\frac{r^2}{1+H\rho'^2}\right)\right] \partial_r \rho_0 + \w^2 (1 + H \r'^2)\, \rho_0=0,
\eeq
while the equation for $A_t^0$ becomes simply $\partial_r$ of eq.~\eqref{eq:rho0eom}. A solution of eq.~\eqref{eq:rho0eom} has an asymptotic expansion around $r=0$ of the form in eq.~\eqref{eq:rhosmallr}
\beq
\rho_0 = \frac{c^{\rho}_{-1}}{r} + d^{\rho}_0 + \mathcal{O}\left(r\right),
\eeq
where $c^{\rho}_{-1}$ and $d^{\rho}_0$ are clearly fluctuations of the parameters $Q$ and $\rhov$ in the background solution eq.~\eqref{basicspike}, respectively. Changing $Q$ means changing the representation of the holographically dual 1/2-BPS Wilson line, hence we will take $c^{\rho}_{-1}=0$ in all that follows. In other words, we will demand that $\rho_0$ is normalisable in the asymptotically $AdS_2$ region $r \to 0$.

In fact, in all that follows we will consider only fluctuations normalisable in the asymptotically $AdS_2$ region $r \to 0$, that is, we will always take $c^{\rho}_{-l-1}=0$ and $c^t_{-l-1}=0$ in eq.~\eqref{eq:eomsmallr} and $c^B_{-l}=0$ in eq.~\eqref{eq:abetasmallr}, respectively.

In sec.~\ref{sec:qnms} we will compute QNM solutions of eqs.~\eqref{eq:leoms} and~\eqref{eq:abetaeom}, and in sec.~\ref{sec:scattering} we will compute scattering solutions. Both types of solutions will be normalisable in the asymptotically $AdS_2 \times S^2$ region $r \to 0$, as described in the previous paragraph. The difference between them will appear in the boundary conditions in the asymptotically Minkowski region $r \to \infty$. For QNMs we will dial through values of $\omega$ until we find normalisable solutions that are purely out-going in the asymptotically Minkowski region, meaning $f^{\rho}_l(r)=0$ and $f^t_l(r)=0$ in eq.~\eqref{eq:eomlarger}, or $f^B_l(r)=0$ in eq.~\eqref{eq:abetalarger}. For scattering solutions we will imagine shooting waves in from infinity and measuring what comes back out at infinity, so we allow both in- and out-going waves in the Minkowski region, but where the out-going waves may have a phase shift compared to the in-going waves. For example in eq.~\eqref{eq:eomlarger} the complex-valued coefficient $g_l^{\rho}(r)$ may have a phase shift compared to $f_l^{\rho}(r)$, and similarly for the other fluctuations.

\section{Quasi-Normal Modes}
\label{sec:qnms}

QNMs are normalisable, out-going solutions for the fluctuations. We compute the QNMs via numerical shooting with parameter $\omega$, the complex frequency. More specifically, we dial through $\omega$ values, for each value numerically solving the fluctuations' equations of motion, always imposing normalisability in the asymptotically $AdS_2 \times S^2$ region $r \to 0$, until we find a purely out-going solution in the asymptotically Minkowski region $r \to \infty$. The details of the fluctuations' asymptotics in both regions, and precisely which coefficients we set to zero to define normalisability and out-going waves, appear in the previous section.

The dimensionless free parameters of the spike are $\rhov/L$, which in the CFT determines $\vev$, and $Q$, which in the CFT determines the number of boxes in the Young tableau of the Wilson line's representation, as described below eq.~\eqref{eq:ads2s2}. Our main question in this section will thus be how the QNM frequencies, in units of $\rhov/L$, vary with $Q$.

\begin{figure}
	\begin{subfigure}{0.5\textwidth}
		\includegraphics[width=\textwidth]{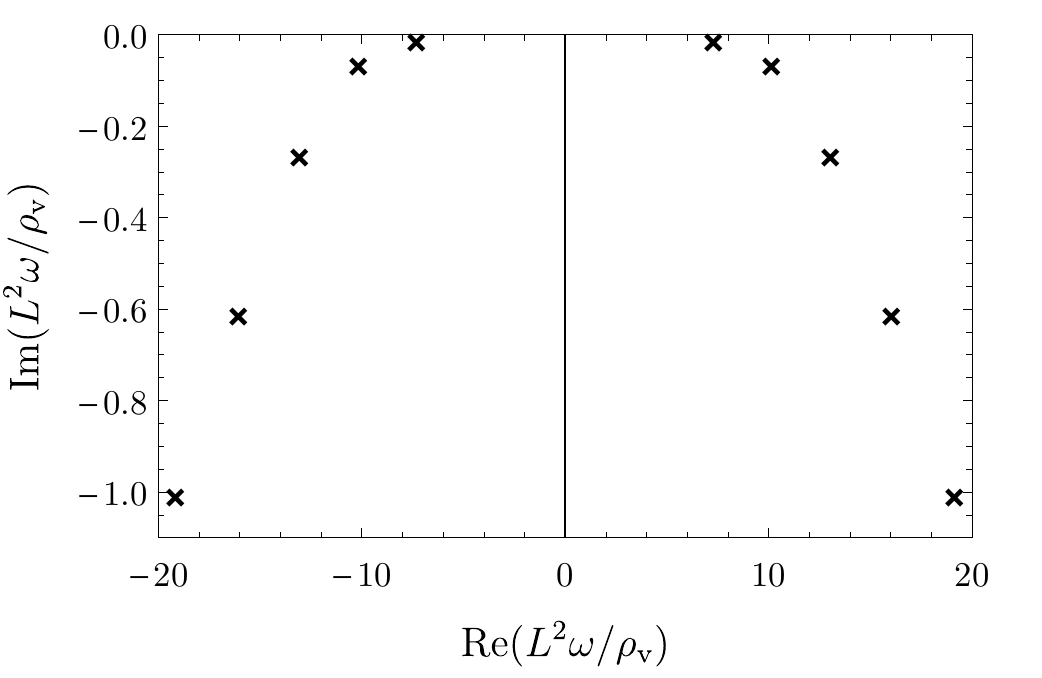}
		\caption{$Q = 0.01$}
	\end{subfigure}
	\begin{subfigure}{0.5\textwidth}
		\includegraphics[width=\textwidth]{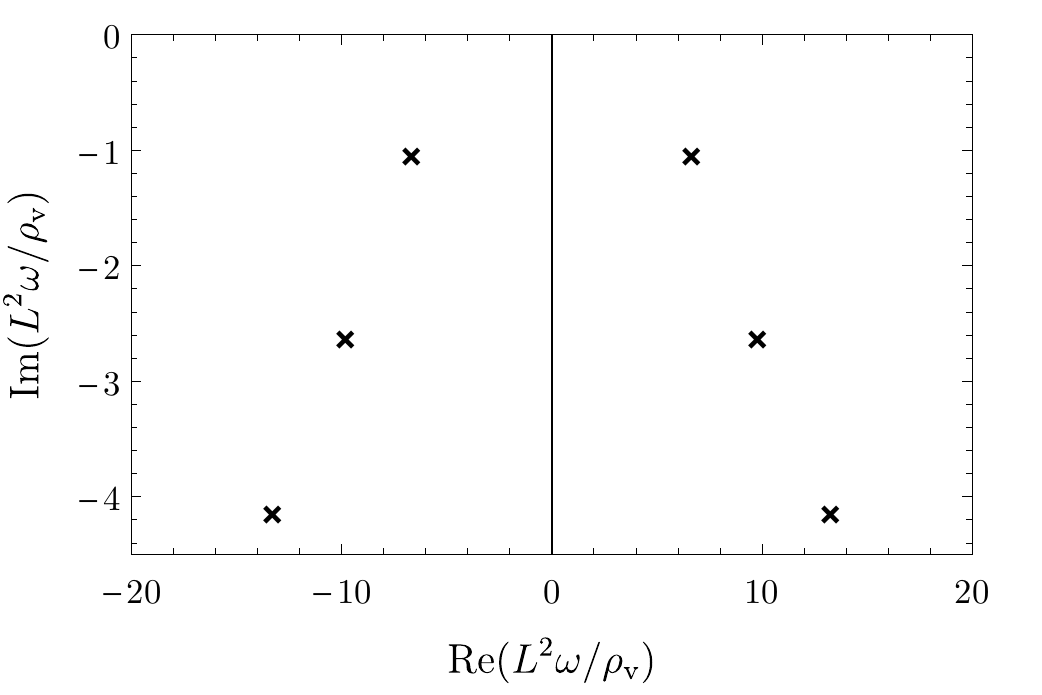}
		\caption{$Q = 0.1$}
	\end{subfigure}
	\caption{The complex frequency plane $\omega$, in units of $\rhov/L^2$. The black crosses denote QNM frequencies of the fluctuation $\rho_0$ when (a) $Q = 0.01$ or (b) $Q = 0.1$.
	}
	\label{fig:s_wave_complex_plane1}
\end{figure}

\begin{figure}
	\begin{center}
		\includegraphics[width=0.6\textwidth]{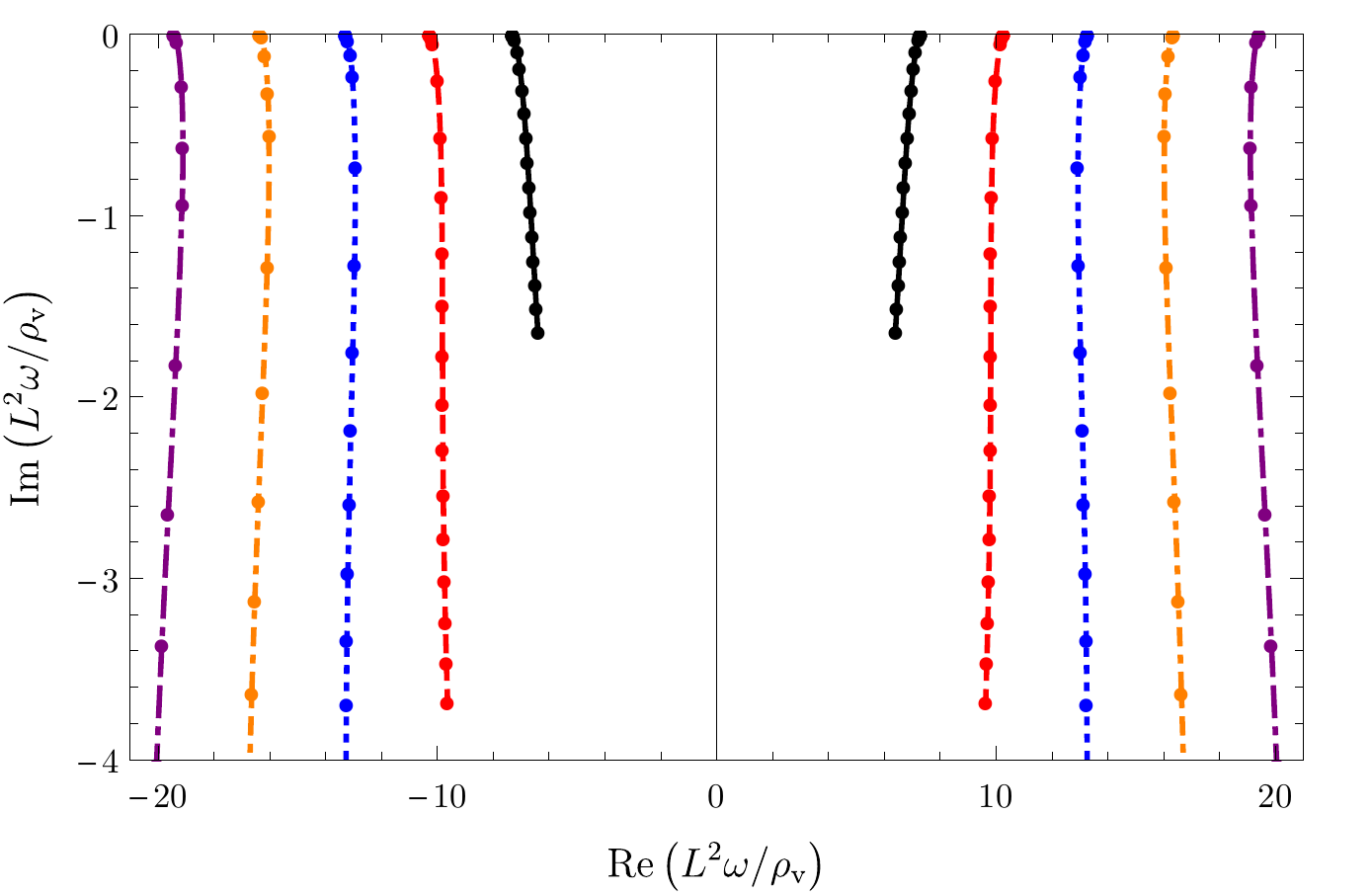}
	\end{center}
	\caption{The complex frequency plane $\omega$, in units of $\rhov/L^2$, showing QNM frequencies of the fluctuation $\rho_0$. The curves (solid black, dashed red, dotted blue, dash-dotted orange, dash-dash-dotted purple) are simply to guide the eye for each QNM's motion down into the complex plane as $Q$ increases from $10^{-5}$ to $10^{-1}$. The curves connect the solid disks, which are separated by 10 data points.}
	\label{fig:s_wave_complex_plane2}
\end{figure}

\begin{figure}
\begin{subfigure}{0.5\textwidth}
    \includegraphics[width=\textwidth]{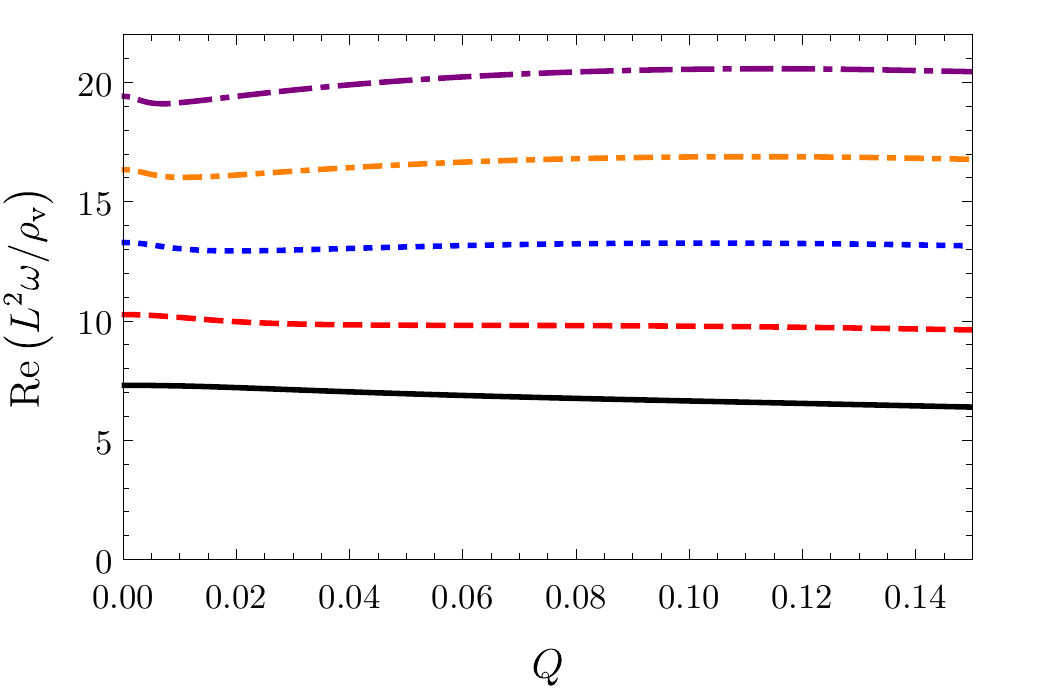}
    \caption{}
\end{subfigure}
\begin{subfigure}{0.5\textwidth}
    \includegraphics[width=\textwidth]{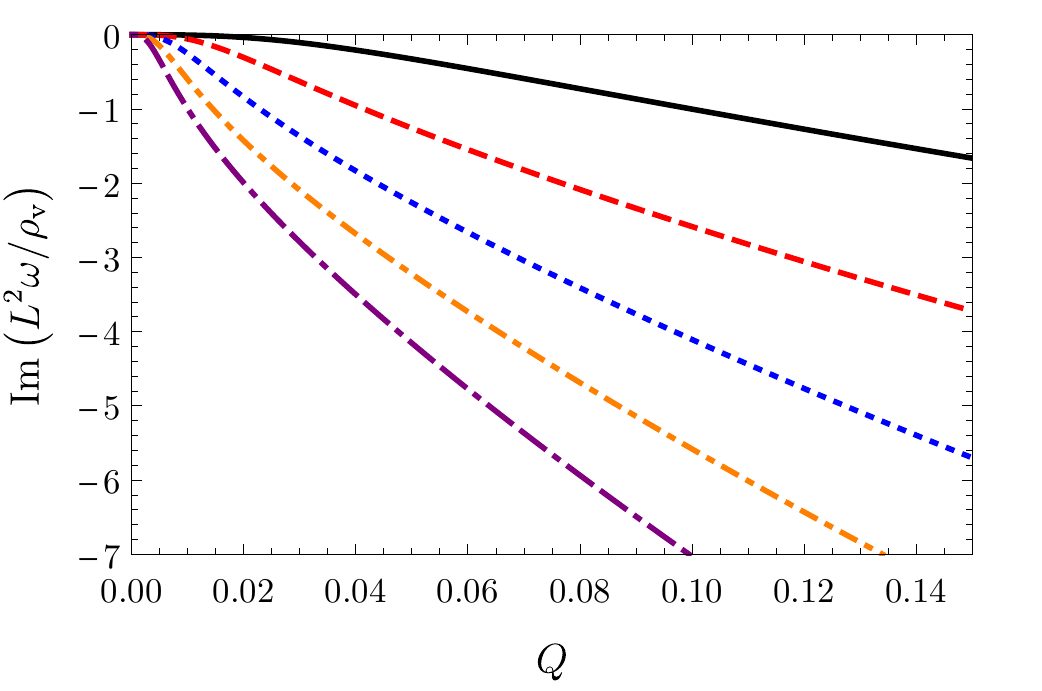}
    \caption{}
\end{subfigure}
\begin{subfigure}{0.5\textwidth}
    \includegraphics[width=\textwidth]{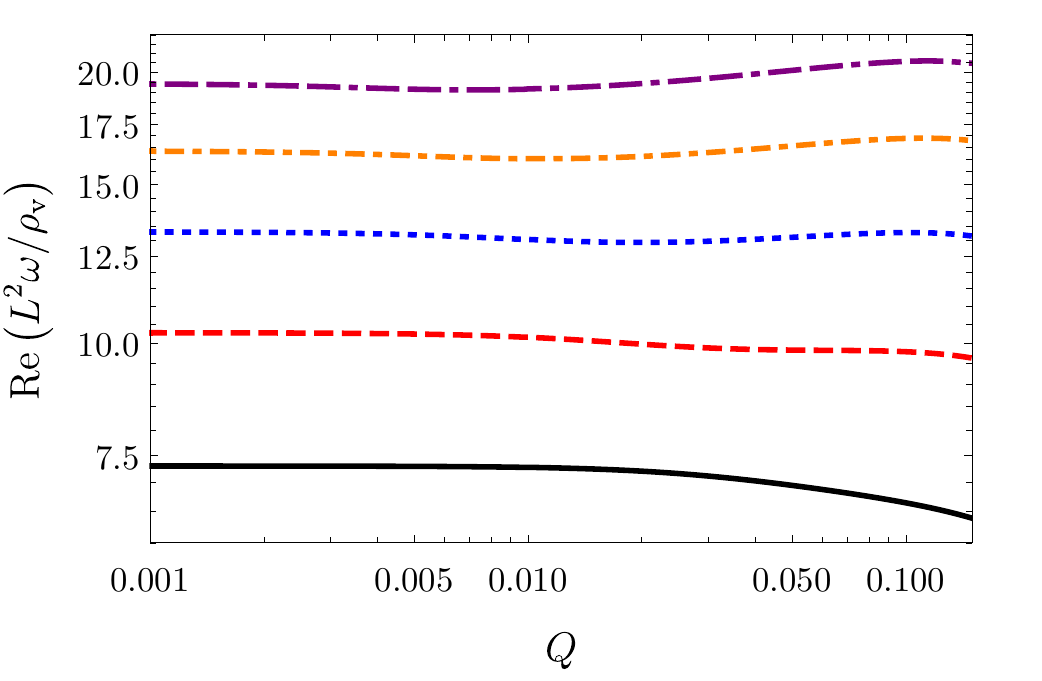}
    \caption{}
\end{subfigure}
\begin{subfigure}{0.5\textwidth}
    \includegraphics[width=\textwidth]{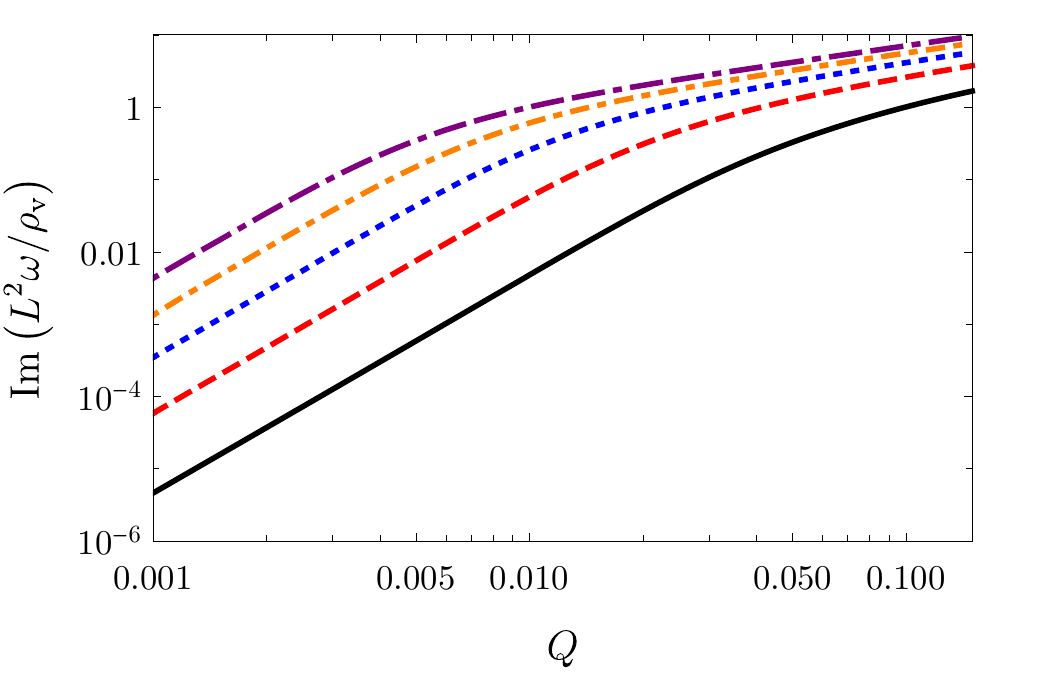}
    \caption{}
\end{subfigure}
\caption{The real and imaginary parts of QNM frequencies of the fluctuation $\rho_0$, in units of $\rhov/L^2$, as functions of $Q$. (a) and (b) have linear axes while (c) and (d) have logarithmic axes. The curves (solid black, dashed red, dotted blue, dash-dotted orange, dash-dash-dotted purple) are to guide the eye, and denote the same QNMs as in fig.~\ref{fig:s_wave_complex_plane2}.}
\label{fig:s_wave_complex_plane3}
\end{figure}

For the fluctuation $\rho_0$, with equation of motion in eq.~\eqref{eq:rho0eom}, our numerical results for the QNM frequencies when $Q=0.01$ and $0.1$ appear in fig.~\ref{fig:s_wave_complex_plane1}. We find the familiar ``Christmas tree" pattern of QNMs typical of holographic systems (though here without a black brane horizon), namely two sets of QNMs reflection-symmetric about the $\textrm{Im}(\omega)$ axis due to time-reversal symmetry, and descending into the complex $\omega$ plane with increasing real parts and increasingly negative imaginary parts. In particular, going from $Q = 0.01$ in fig.~\ref{fig:s_wave_complex_plane1} (a) to $Q=0.1$ in fig.~\ref{fig:s_wave_complex_plane1} (b), the QNMs' real parts change very little, while the imaginary parts become more negative, or in physical terms, the QNMs become less stable. Figs.~\ref{fig:s_wave_complex_plane2} and~\ref{fig:s_wave_complex_plane3} show the same phenomenon in more detail, for other ranges of $Q$. One possible interpretation of this phenomenon is that in the CFT larger $Q$, and hence a larger Young tableau, provides more ``decay channels'' for excitations at the impurity. Fig.~\ref{fig:s_wave_complex_plane3} (d) is a log-log plot of $\textrm{Im}\left(\omega\right)$ versus $Q$, and reveals a transition between power laws as $Q$ increases, from $\textrm{Im}\left(\omega\right) \propto Q^3$ at small $Q$ to $\textrm{Im}\left(\omega\right) \propto Q^{3/4}$ at large $Q$.

\begin{figure}
\begin{subfigure}{0.5\textwidth}
    \includegraphics[width=\textwidth]{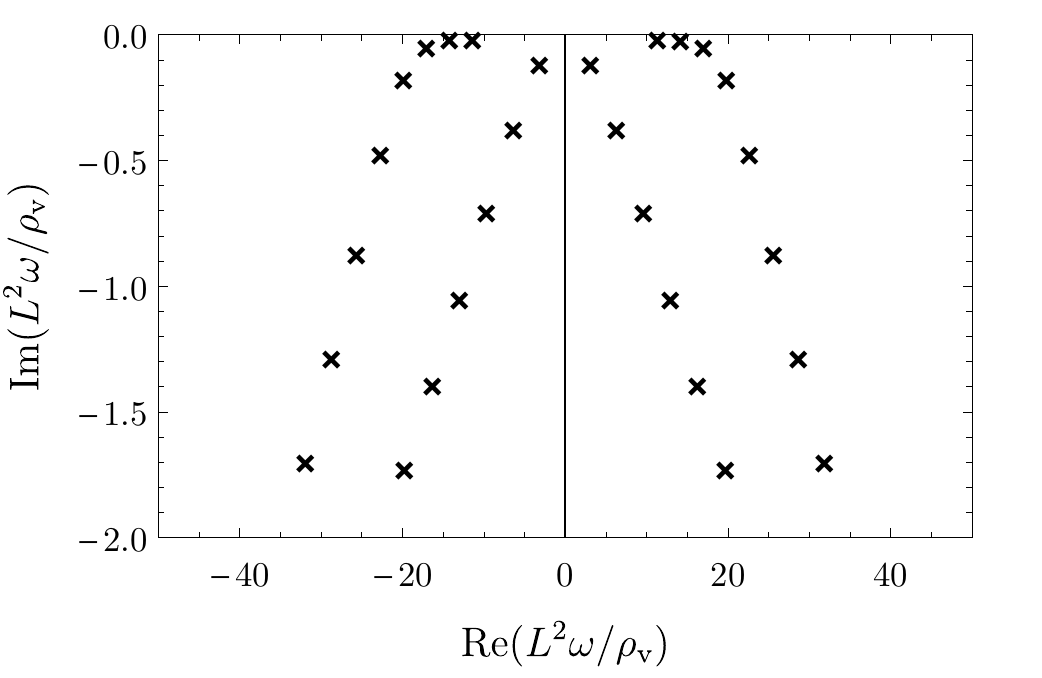}
    \caption{ $l = 1$ and $Q = 0.01$}
\end{subfigure}
\begin{subfigure}{0.5\textwidth}
    \includegraphics[width=\textwidth]{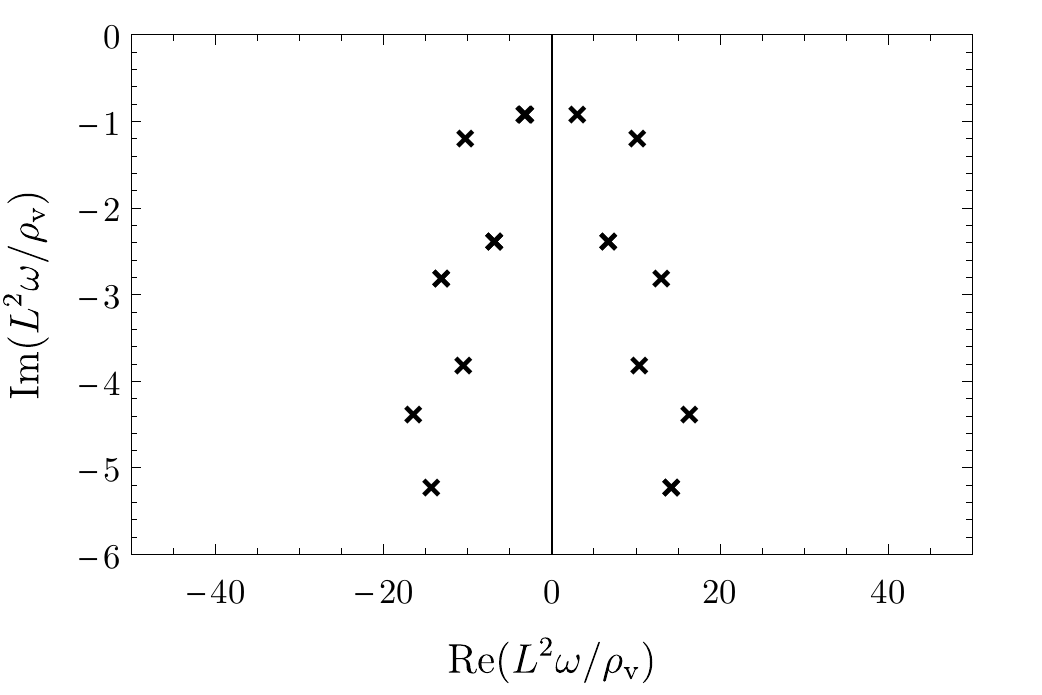}
    \caption{$l = 1$ and $Q = 0.1$}
\end{subfigure}
\begin{subfigure}{0.5\textwidth}
    \includegraphics[width=\textwidth]{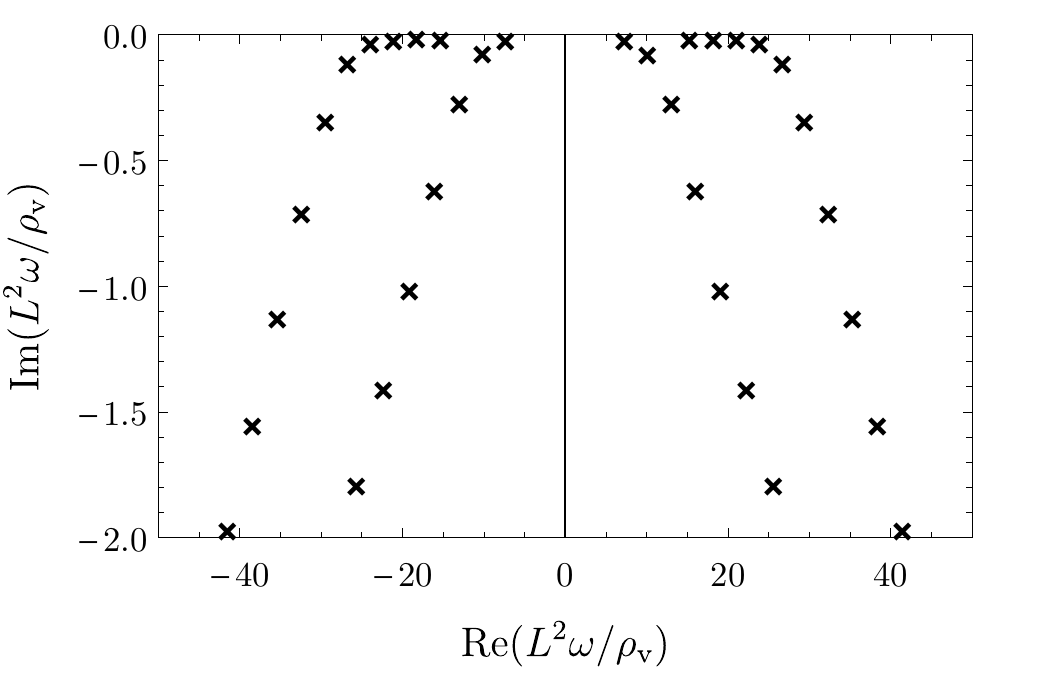}
    \caption{$l = 2$ and $Q = 0.01$}
\end{subfigure}
\begin{subfigure}{0.5\textwidth}
    \includegraphics[width=\textwidth]{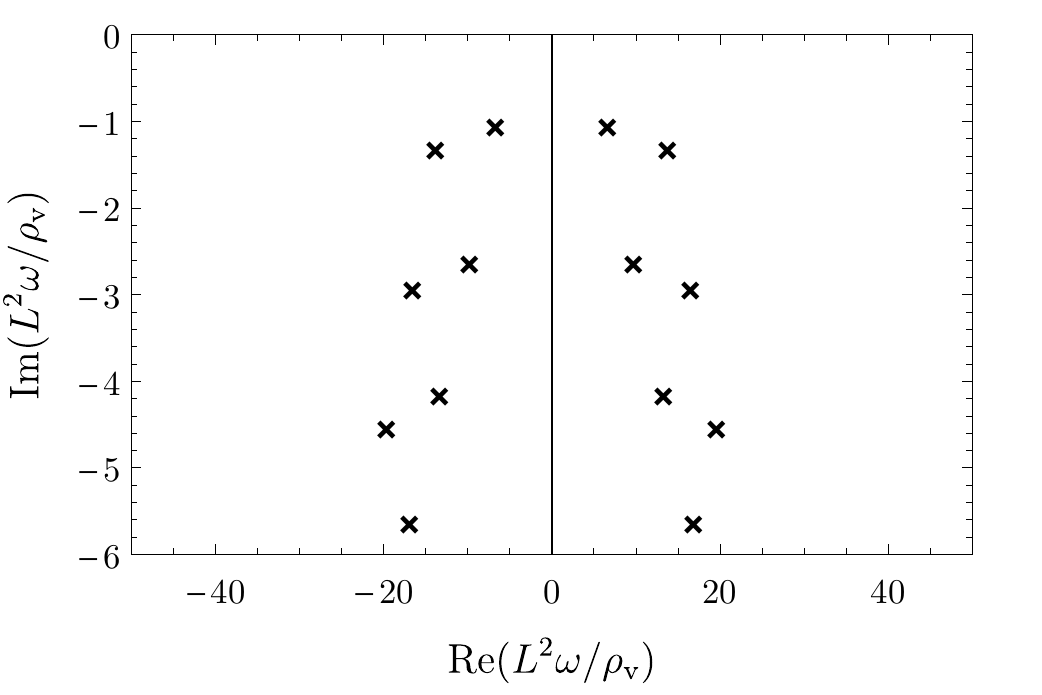}
    \caption{$l = 2$ and $Q = 0.1$}
\end{subfigure}
\begin{subfigure}{0.5\textwidth}
    \includegraphics[width=\textwidth]{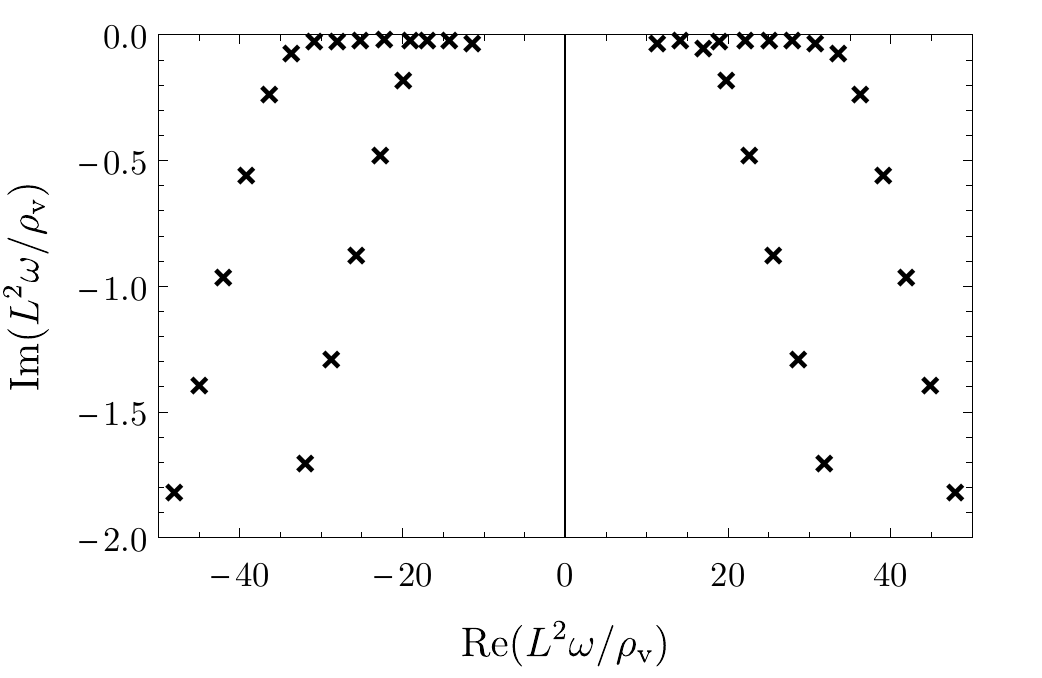}
    \caption{$l = 3$ and $Q = 0.01$}
\end{subfigure}
\begin{subfigure}{0.5\textwidth}
    \includegraphics[width=\textwidth]{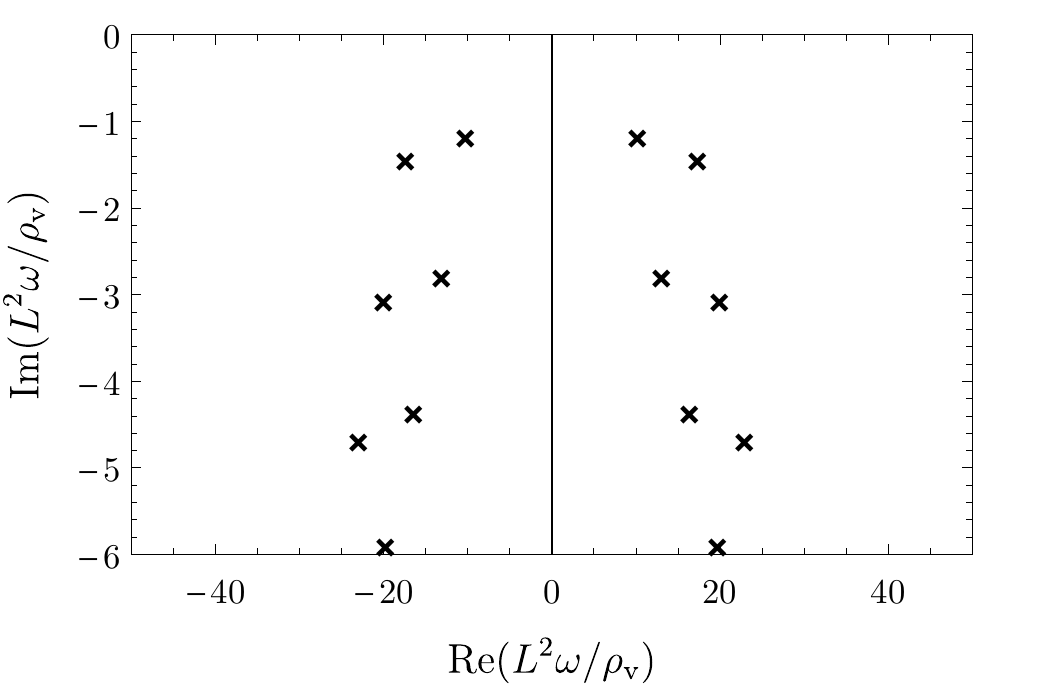}
    \caption{$l = 3$ and $Q = 0.1$}
\end{subfigure}
\caption{The complex frequency plane $\omega$, in units of $\rhov/L^2$. The black crosses denote QNM frequencies of the fluctuations $\rho_l$ and $A_t^l$ when $l=1$ (top row), $l=2$ (middle row), and $l=3$ (bottom row), for $Q = 0.01$ (left column) and $Q = 0.1$ (right column).
}
\label{fig:higher_wave_complex_plane}
\end{figure}

Fig.~\ref{fig:higher_wave_complex_plane} shows our numerical results for the QNMs of the fluctuations $\rho_l$ and $A_t^l$, with equations of motion in eq.~\eqref{eq:leoms}, when $l=1,2,3$ and $Q=0.01$ and $0.1$. We find again that if we fix $l$ and increase $Q$ (moving from left to right in one row of fig.~\ref{fig:higher_wave_complex_plane}) then the QNMs' real parts change very little while the imaginary parts become more negative, or in other words the QNMs become less stable.

If we instead fix $Q$ and increase $l$ (moving down one column of fig.~\ref{fig:higher_wave_complex_plane}), then we find two different behaviours, depending on the size of $Q$. If $Q$ is sufficiently small, then as we increase $l$ (moving down the left column of fig.~\ref{fig:higher_wave_complex_plane}) the QNMs tend to move up, closer to the $\textrm{Re}(\omega)$ axis, thus becoming more stable. In fact, as $l$ increases, more and more QNMs ``line up'' just below the $\textrm{Re}(\omega)$ axis. If we define the number of QNMs lining up this way as QNMs with $|\textrm{Im}\left(\omega\right)|/\textrm{Re}\left(\omega\right)<10^{-2}$, then the number appears to grow approximately linearly with $l$. Our numerics also suggest that this line of QNMs approaches the $\textrm{Re}\left(\omega\right)$ axis exponentially quickly in $l$ as $l \to \infty$. They may thus appear to be forming a branch cut on the real axis as $l$ increases, however they actually maintain order one spacing from each other as $l \to \infty$, in units of $\rhov/L^2$, and in fact are not evenly spaced, or indeed have any order in their spacing that we could discern. On the other hand, if $Q$ is sufficiently large, then as we increase $l$ (moving down the right column of fig.~\ref{fig:higher_wave_complex_plane}) the QNMs' real parts grow larger while the imaginary parts become more negative, or in other words the QNMs become less stable. The critical value of $Q$ that separates the two behaviours is $Q\approx 0.078$.

\begin{figure}
    \begin{subfigure}{\textwidth}
        \begin{center}
            \includegraphics{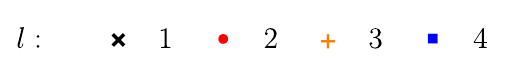}
            \vspace{-0.25cm}
        \end{center}
    \end{subfigure}
    \begin{subfigure}{0.5\textwidth}
        \begin{center}
        \includegraphics[width=\textwidth]{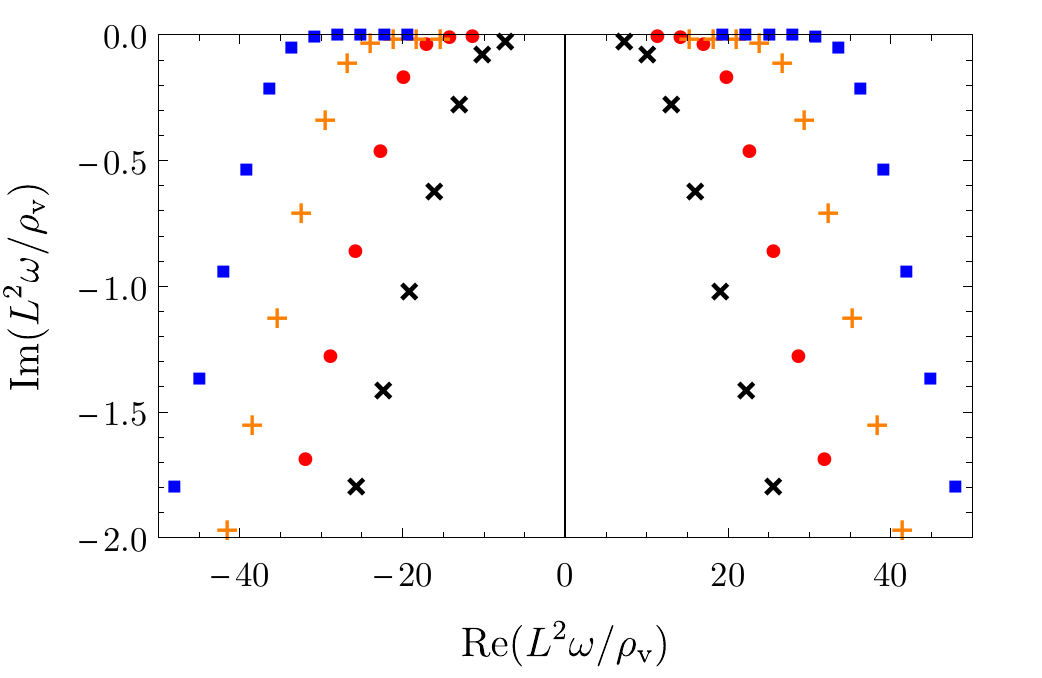}
        \caption{\(Q = 0.01\)}        
        \end{center}
    \end{subfigure}
    \begin{subfigure}{0.5\textwidth}
        \begin{center}
        \includegraphics[width=\textwidth]{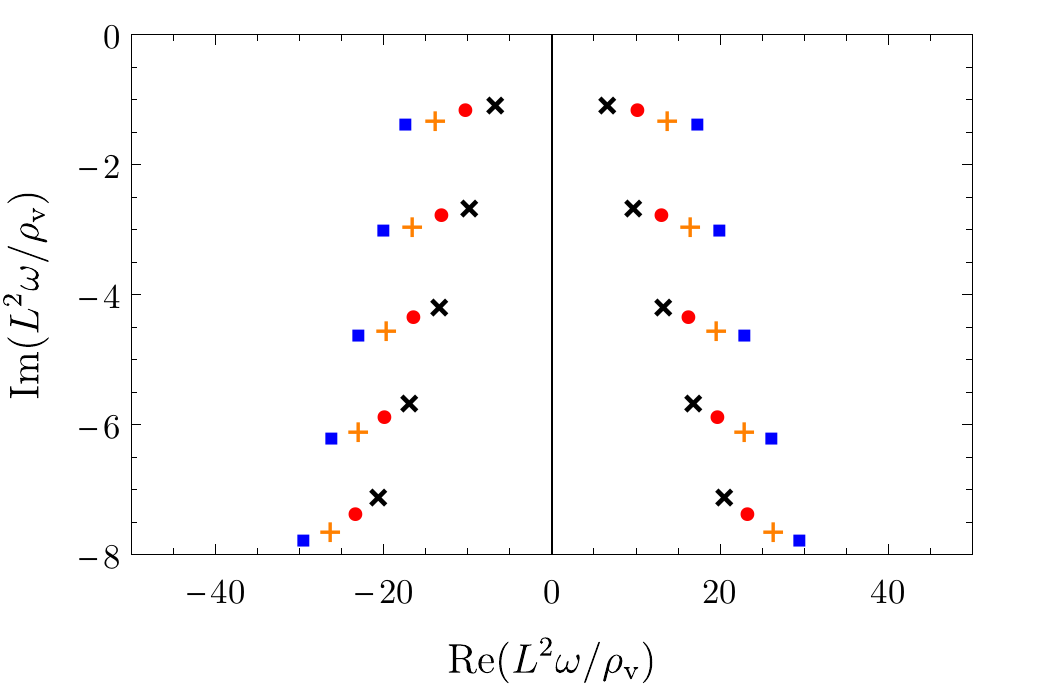}
        \caption{\(Q = 0.1\)}        
        \end{center}
    \end{subfigure}
    \caption{The complex frequency plane $\omega$, in units of $\rhov/L^2$, showing QNM frequencies of the fluctuations $B_l$ for (a) $Q = 0.01$ or (b) $Q = 0.1$ and $l=1,2,3,4$ (black crosses, red dots, orange plus signs, blue squares, respectively). }
    \label{fig:qnm_beta_multiple_l}
\end{figure}

For the vector harmonic fluctuations $B_l$, with equation of motion in eq.~\eqref{eq:abetaeom}, our numerical results for the QNM frequencies when $l=1,2,3,4$ and $Q = 0.01$ and $0.1$ appear in fig.~\ref{fig:qnm_beta_multiple_l}. We find similar behaviour to the fluctuations of $\rho_l$ and $A_t^l$ in figs.~\ref{fig:s_wave_complex_plane1} and~\ref{fig:higher_wave_complex_plane}. In particular, we find again that if we fix $l$ and increase $Q$ (moving from fig.~\ref{fig:qnm_beta_multiple_l} (a) to (b)) then the QNMs' real parts change very little but the imaginary parts become more negative, i.e. the QNMs become less stable. If instead we fix $Q$ and increase $l$ then again we find for small $Q$ (fig.~\ref{fig:qnm_beta_multiple_l} (a)) the QNMs' imaginary parts become less negative, i.e. the QNMs become more stable, and some QNMs line up just below the $\textrm{Re}\left(\omega\right)$ axis, with the same properties as before: as $l$ increases, the number of QNMs that line up grows approximately linearly in $l$, the line of QNMs approaches the $\textrm{Re}\left(\omega\right)$ axis exponentially quickly in $l$, and the QNMs in the line maintain an uneven but order-one spacing from one another. On the other hand, for large $Q$ (fig.~\ref{fig:qnm_beta_multiple_l} (b)), as $l$ increases the QNMs' imaginary parts become more negative, i.e. the QNMs become less stable. The critical value of $Q$ that separates the two behaviours is again $Q \approx 0.078$.

We can gain useful intuition for these features of the QNMs by re-writing each fluctuation's equation of motion in the form of a Schr\"odinger equation. The equations of motion for $\rho_0$ in eq.~\eqref{eq:rho0eom} and for $B_l$ in eq.~\eqref{eq:abetaeom} are of the generic form
\begin{equation}
\label{eq:y}
    \partial_r^2 y(r) + \g_1(r) \partial_r y(r) + \le[ \w^2 \g_2(r) - \frac{l(l+1)}{r^2} \ri]  y(r) = 0,
\end{equation}
where $y(r)$ denotes the fluctuation $\rho_0$ or $B_l$, and the coefficients $\g_1(r)$ and $\g_2(r)$ both depend on $\rhov/L$ and $Q$ but not on $\omega$. If we define a new radial coordinate $\zeta$ via $\partial_r \zeta \equiv \sqrt{\g_2(r)}$, with the boundary condition that $\zeta = 0$ when $r=0$, and we define a new fluctuation $\Psi$ via
\beq
\Psi  \equiv \zeta \,e^h \, y,
\eeq
where $h$ is a function of $\zeta$ determined by
\beq
\partial_{\zeta} h \equiv -\frac{1}{\zeta} + \frac{\partial_r \g_2(r) + 2 \g_1(r)\g_2(r)}{4 \, \g_2(r)^{3/2}},
\eeq
where of course $r$ is implicitly a function of $\zeta$, then $y(r)$'s equation of motion eq.~\eqref{eq:y} becomes a Schr\"odinger equation for $\Psi(\zeta)$,
\beq
\partial_{\zeta}^2 \Psi + \left [\omega^2 - \frac{l(l+1)}{\zeta^2} - V(\zeta) \right] \Psi= 0,
\eeq
with potential $V(\zeta)$ determined by $\g_1(r)$, $\g_2(r)$, and $l$,
\beq
\label{eq:schrodinger_potential}
V(\zeta) \equiv \frac{1}{4}\frac{\partial_r^2\g_2(r)}{\g_2(r)^2} - \frac{5}{16}\frac{\left(\partial_r \g_2(r)\right)^2 }{\g_2(r)^3}+ \frac{2 \partial_r\g_1(r) + \g_1(r)^2}{4 \g_2(r)} + l(l+1) \le[ \frac{1}{r^2 \g_2(r)} - \frac{1}{\zeta^2} \ri],
\eeq
where again $r$ is implicitly a function of $\zeta$.

\begin{figure}
    \begin{center}
	\begin{subfigure}{\textwidth}
		\begin{center}
		\includegraphics[width=0.67\textwidth]{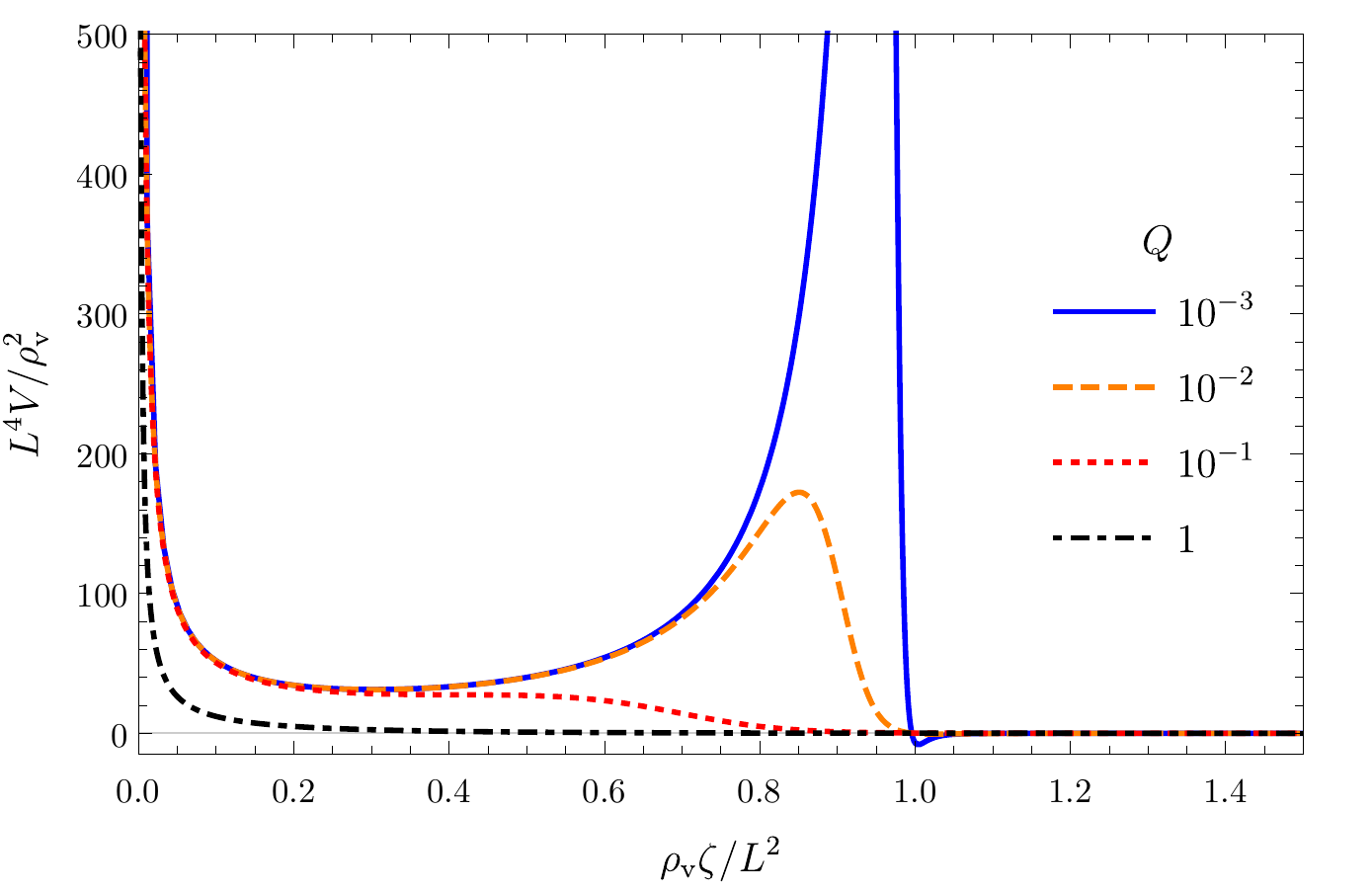}
		\end{center}
		\caption{}
		\label{fig:schrodinger_potential_main}
    \end{subfigure}
   \end{center}
    \begin{subfigure}{0.5\textwidth}
        \includegraphics[width=\textwidth]{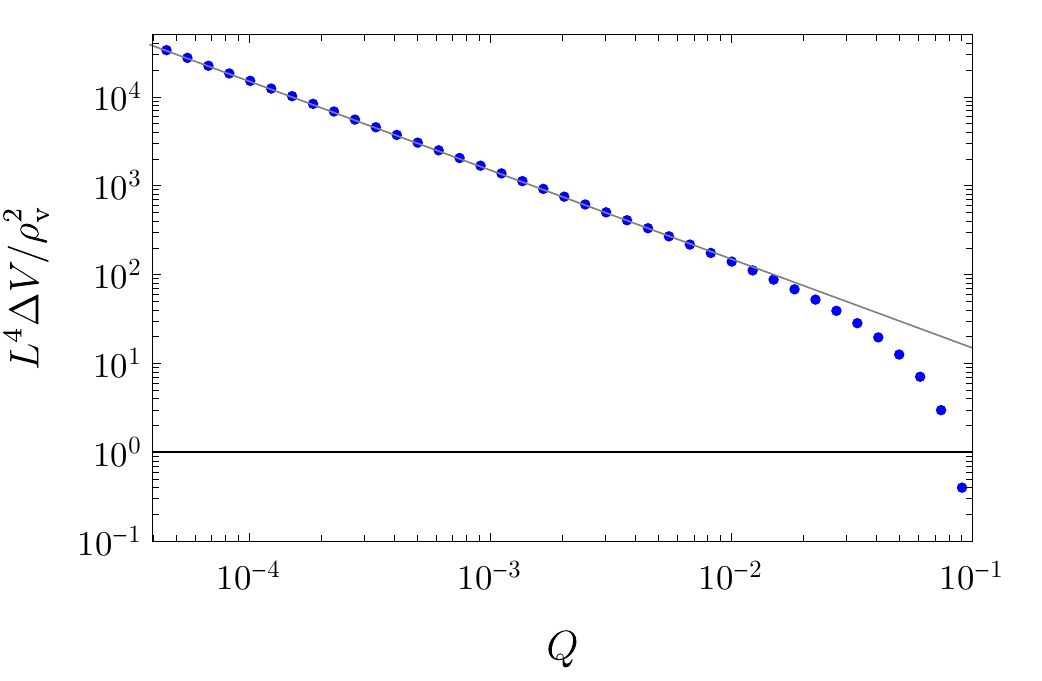}
        \caption{}
    \end{subfigure}
    \begin{subfigure}{0.5\textwidth}
        \includegraphics[width=\textwidth]{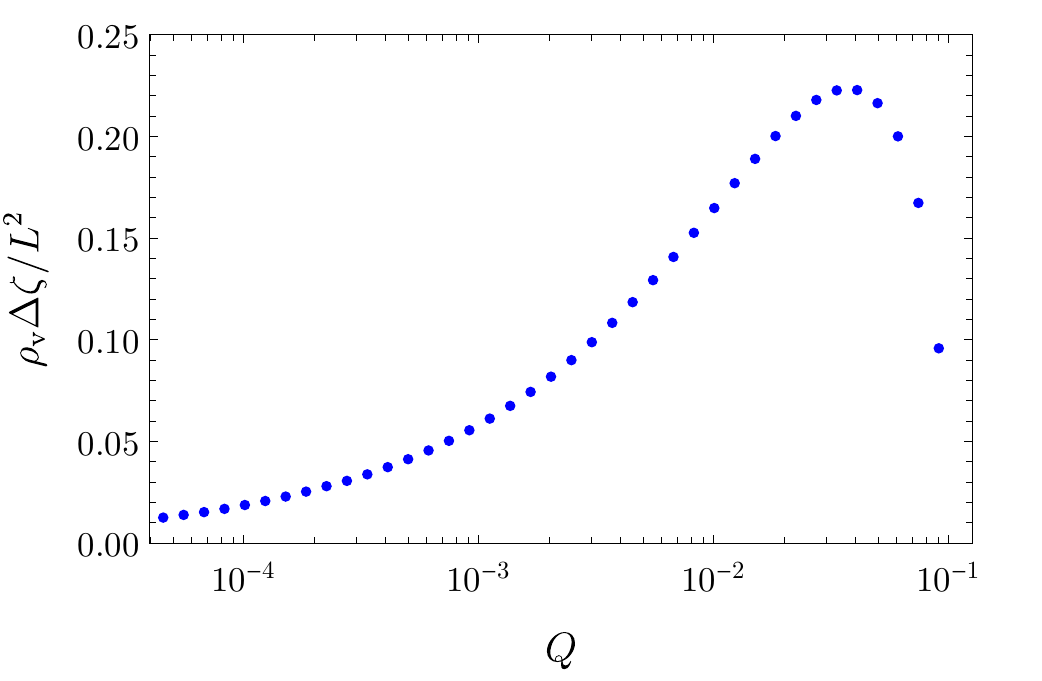}
        \caption{}
    \end{subfigure}
    \caption{(a) The Schr\"odinger potential $V(\zeta)$ defined in eq.~\eqref{eq:schrodinger_potential}, in units of $\rhov^2/L^4$, as a function of $\zeta$, in units of $L^2/\rhov$, for the fluctuation $\rho_0$ with $Q=10^{-3},10^{-2},10^{-1},1$ (solid blue, dashed orange, dotted red, dash-dotted black). The asymptotic behaviours are given in eq.~\eqref{eq:rho0potentialasymptotics}: as $\zeta \to 0$ an infinite barrier appears, while $V \to 0^-$ as $\zeta \to \infty$. For small $Q$ a peak appears at finite $\zeta$, and hence a potential well, with local minimum at $V>0$. As $Q$ increases, the peak shrinks and broadens, and eventually disappears when $Q \approx 0.0977$. (b) Logarithmic plot of the peak's height, $\Delta V$, defined as the difference in $V$ between the top of the peak and the bottom of the potential well, in units of $\rhov^2/L^4$, as a function of $Q$. The grey line is $\Delta V = 3 \rhov^2/(2 Q L^4)$, which is a good approximation for $Q\lesssim 10^{-2}$. (c) The full width at half maximum of the peak, $\Delta \zeta$, in units of $L^2/\rhov$, as a function of $Q$.}
     \label{fig:schrodinger_potential}
\end{figure}

For $\rho_0$, the equation of motion in eq.~\eqref{eq:rho0eom} has
\beq
    \g_1(r) = \frac{2}{r} + \frac{4 Q^2 \rhov}{(Q L^2 + \rhov \, r) \le[Q^2 + (Q + \rhov \, r/L^2)^4 \ri]},
    \qquad
    \g_2(r) = 1 + \frac{Q^2}{\left(Q + \rhov \, r /L^2\right)^4}.
\eeq
From $\zeta$'s definition $\partial_r \zeta \equiv \sqrt{\g_2(r)}$, with the boundary condition that $\zeta = 0$ when $r=0$, we find that $\zeta$ is given by
\beq
\zeta = \frac{1}{\rhov} \left [ \left(Q L^2 + \rhov \, r\right){}_2F_1\left(-\frac{1}{2},-\frac{1}{4};\frac{3}{4};-\frac{Q^2}{\left(Q + \rhov\,r / L^2\right)^4}\right)-Q L^2 \, {}_2F_1\left(-\frac{1}{2},-\frac{1}{4};\frac{3}{4};-\frac{1}{Q^2} \right)\right],
\label{eq:zeta}
\eeq
with leading-order asymptotics
\beq
\zeta = \begin{cases} r\,\left(1 + Q^{-2}\right)^{1/2},  & r \to 0,\\ r,  & r \to \infty. \end{cases}
\eeq
The Schr\"odinger potential obtained from eq.~\eqref{eq:schrodinger_potential} then has leading-order asymptotics
\beq
\label{eq:rho0potentialasymptotics}
V(\zeta) = \begin{cases} \frac{4 \rhov}{(1 + Q^2)^{3/2} L^2}\,\zeta^{-1},  & \zeta \to 0,\\  & \\ -\frac{Q^2 L^8}{\rhov^4}\,\zeta^{-6},  & \zeta \to \infty. \end{cases}
\eeq
For any $Q$ this $V(\zeta)$ approaches $+\infty$ as $\zeta \to 0$ and approaches zero from below as $\zeta \to +\infty$, and hence has a global minimum with $V<0$ at some finite $\zeta$. Fig.~\ref{fig:schrodinger_potential} (a) shows our numerical results for this $V(\zeta)$ for several $Q$ values, showing the expected behaviours as $\zeta\to0$ and $\zeta \to \infty$. In fig.~\ref{fig:schrodinger_potential} (a) the global minimum with $V<0$ at finite $\zeta$ is only visible for small $Q$, but is indeed present for all $Q$, as we have confirmed numerically.

However, another key feature of $V(\zeta)$ in fig.~\ref{fig:schrodinger_potential} (a) is, for sufficently small $Q$, a peak of finite height at some finite $\zeta$, producing a potential well with a minimum at $V>0$. As $Q$ increases, this peak becomes shorter and wider, although the former occurs more rapidly than the latter: comparing $V$ for $Q = 10^{-3}$ and $10^{-2}$ in fig.~\ref{fig:schrodinger_potential}, clearly the peak is smaller when $10^{-2}$. As $Q$ increases beyond $Q \approx 0.0977$ the peak becomes so short and wide as to disappear completely. Our numerical results for the peak's height $\Delta V$, defined as the difference in $V$ between the top of the peak and the bottom of the potential well (i.e. between the local maximum and local minimum), as a function of $Q$ appears in fig.~\ref{fig:schrodinger_potential} (b). For $Q \lesssim 10^{-2}$ we find that $\Delta V$ is well-approximated by $3 \rhov^2/(2QL^4)$. Our numerical results for the peak's full width at half maximum, $\Delta \zeta$, as a function of $Q$ appears in fig.~\ref{fig:schrodinger_potential} (c).

Some solutions of the Schr\"odinger equation will be ``bound" in the potential well between the infinite barrier on the left at $\zeta=0$ and the finite barrier provided by the peak on the right. More precisely, they will be ``quasi-bound,'' being able to tunnel quantum mechanically under the peak on the right and escape to $\zeta \to \infty$. These solutions will correspond to long-lived QNMs, meaning QNM frequencies with real parts larger than their imaginary parts. For example, when $Q=10^{-2}$ these are the three highest pairs of QNMs in fig.~\ref{fig:s_wave_complex_plane1} (a):  in that case the peak in fig.~\ref{fig:schrodinger_potential} (a) (dashed orange) has a height $ L^4 V/\rhov^2\approx 172$, while the three highest pairs of QNMs in fig.~\ref{fig:s_wave_complex_plane1} (a) have $\left(\textrm{Re}\left(L^2\omega/\rhov\right)\right)^2\approx 53, 103, 170 < 172$.

In contrast, the other QNMs correspond to solutions ``above'' the peak, or present when the peak is absent. Due to the infinite barrier at $\zeta \to 0$ these solutions will always escape to $\zeta \to \infty$, however they have little or nothing in their way, so they will not be long-lived: they correspond to QNM frequencies whose real and imaginary parts are of the same order.

For the vector harmonic fluctuation $B_l$, the equation of motion in eq.~\eqref{eq:abetaeom} has
\beq
\g_1(r) = 0, \qquad \g_2(r) =1 + \frac{Q^2}{\left(Q + \rhov \, r /L^2\right)^4}.
\eeq
Since $\g_2(r)$ is the same as in the \(\rho_0\) equation of motion, \(\zeta\) is again given by eq.~\eqref{eq:zeta}. However, because $\gamma_1(r)$ is different, the asymptotics of the potential are different,
\beq
\label{eq:betapotentialasymptotics}
V(\zeta) = \begin{cases} \frac{2l(l+1)\rhov}{(1 + Q^2)^{3/2} L^2}\,\zeta^{-1},  & \zeta \to 0,\\  & \\ \frac{2l(l+1)Q L^2}{\rhov}\left[1-{}_2F_1\left(-\frac{1}{2},-\frac{1}{4};\frac{3}{4};-\frac{1}{Q^2} \right)\right]\,\zeta^{-3},  & \zeta \to \infty. \end{cases}
\eeq
For any $Q$ this $V(\zeta)$ approaches $+\infty$ as $\zeta \to 0$. In the second line of eq.~\eqref{eq:betapotentialasymptotics} the factor in square brackets is positive for all $Q$, hence for any $Q$ this $V(\zeta)$ approaches zero from above as $\zeta \to \infty$. As a result, nothing requires $V(\zeta)$ to have a global minimum, in contrast to the Schr\"odinger potential associated with $\rho_0$.

\begin{figure}
\begin{subfigure}{\textwidth}
    \begin{center}
        \includegraphics{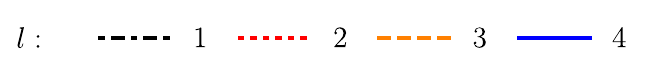}
        \vspace{-0.25cm}
    \end{center}
\end{subfigure}
\begin{subfigure}{0.5\textwidth}
    \includegraphics[width=\textwidth]{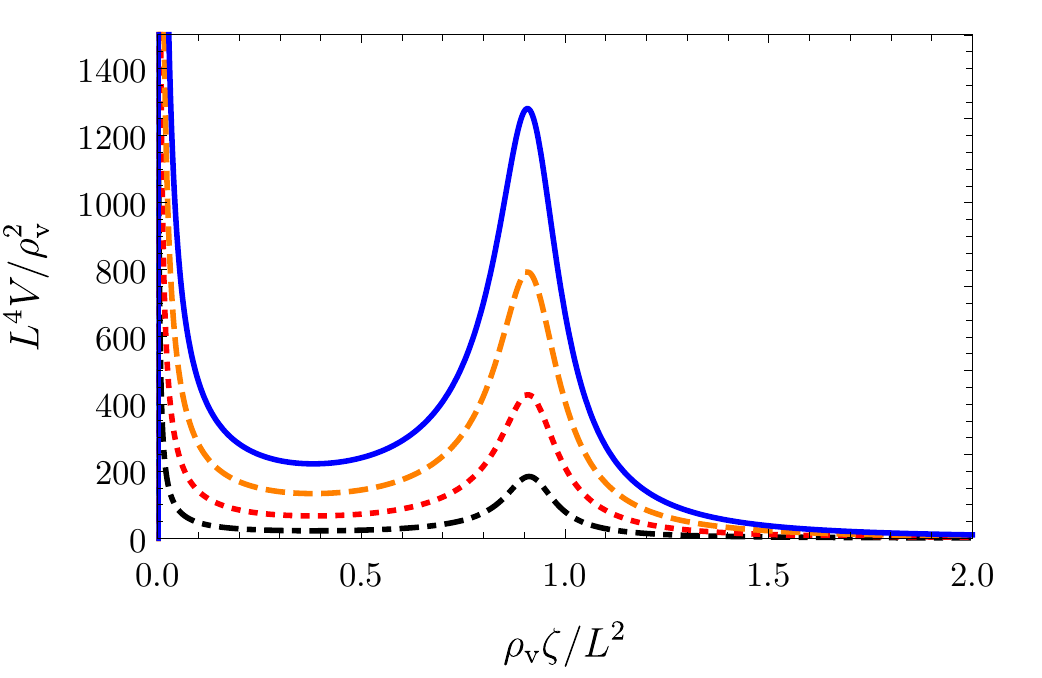}
    \caption{$Q = 0.01$}
\end{subfigure}
\begin{subfigure}{0.5\textwidth}
    \includegraphics[width=\textwidth]{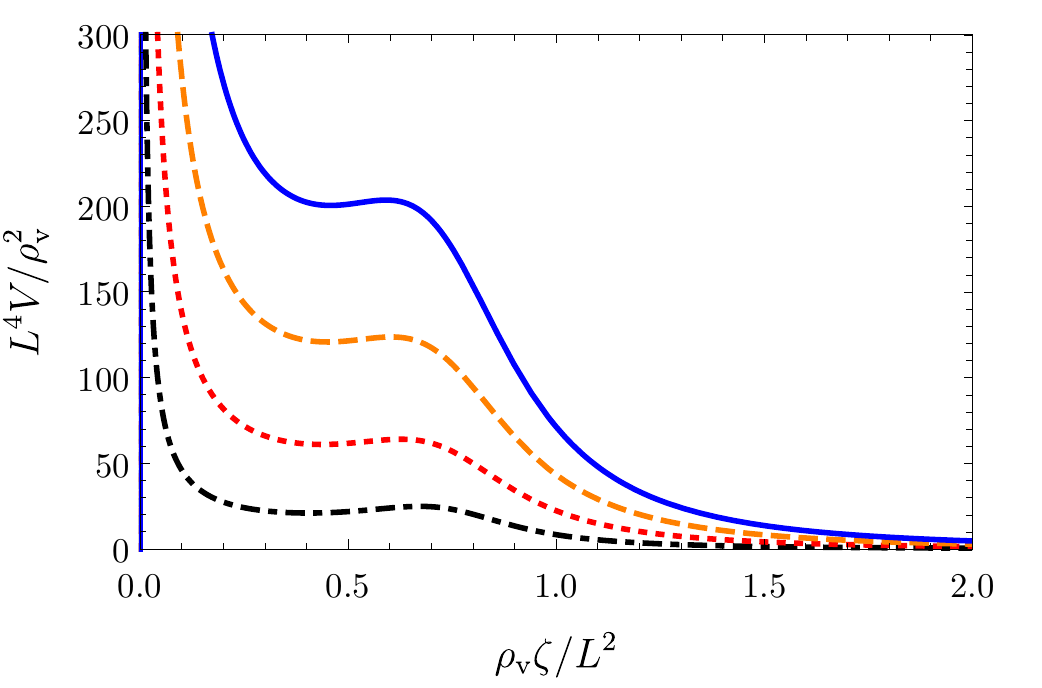}
    \caption{$Q=0.1$}
\end{subfigure}
\caption{The Schr\"odinger potential $V(\zeta)$ defined in eq.~\eqref{eq:schrodinger_potential}, in units of $\rhov^2/L^4$, as a function of $\zeta$, in units of $L^2/\rhov$, for the fluctuation $B_l$ with (a) $Q = 0.01$ and (b) $Q=0.1$ and $l=1,2,3,4$ (black dot-dashed, red dotted, orange dashed, solid blue, respectively).}
\label{fig:beta_potential}
\end{figure}

Fig.~\ref{fig:beta_potential} shows our numerical results for the $V(\zeta)$ of $B_l$ for $Q = 0.01$ and $0.1$ and for $l=1,2,3,4$. We see not only the expected asymptotic behaviours at $\zeta \to 0$ and $\zeta \to \infty$, including in particular an infinite barrier at $\zeta \to 0$, but also a peak at finite $\zeta$, producing a potential well. In this case the peak is present for all $Q$ we considered.

The behaviour of $V(\zeta)$ in fig.~\ref{fig:beta_potential} provides intuition for the behaviour of the QNMs of $B_l$ that we observed in fig.~\ref{fig:qnm_beta_multiple_l}. For example, if we fix $Q = 10^{-2}$ and increase $l$, fig.~\ref{fig:beta_potential} (a) shows the peak growing in height, and correspondingly the potential well growing deeper. We then expect to see more long-lived QNMs: these are the QNMs that ``line up'' below the real axis in fig.~\ref{fig:qnm_beta_multiple_l} (a). If instead we fix $Q = 0.1$, which is above the critical value $Q \approx 0.078$, and increase $l$ then fig.~\ref{fig:beta_potential} (b) shows that the peaks are shorter than at $Q = 10^{-2}$ (compare the vertical axes in figs.~\ref{fig:qnm_beta_multiple_l} (a) and (b)), and although the peak's height grows as $l$ increases, the infinite barrier at $\zeta \to 0$ grows more quickly, so that the potential well actually becomes shallower. As a result, we expect less stable QNMs, as indeed we saw in fig.~\ref{fig:qnm_beta_multiple_l} (b), where the QNM frequencies' imaginary parts become more negative as $l$ increases. If we fix $l$ and increase $Q$, i.e. if we compare the same line between figs.~\ref{fig:qnm_beta_multiple_l} (a) and (b), then we see the potential well become shallower, so in general we expect QNMs to be less stable, as indeed we saw when comparing the same value of $l$ between figs.~\ref{fig:qnm_beta_multiple_l} (a) and (b).

In short, writing the fluctuations' equations of motion in the form of Schr\"odinger equations provides simple but powerful intuition for the behavior of the QNMs. In particular, the Schr\"odinger potentials characteristically exhibit an infinite barrier at the impurity, $r \to 0$, and a finite peak at some finite $r$, producing a potential well. The most stable QNMs correspond to quasi-bound solutions in that potential well. The position of the peak could also potentially provide a definition of the size of the screening cloud surrounding the impurity.

\section{Scattering}
\label{sec:scattering}

In LFLs, screened impurities give rise not only to quasi-bound states, but also to phase shifts in the electronic wave function. Heuristically, if we ``shoot in'' electrons from infinity, they can be trapped at the impurity temporarily before escaping back to infinity. Such resonant scattering produces peaks in the associated scattering cross section.

In this section we will consider scattering off the D3-brane spike. Heuristically, we will ``shoot in'' fluctuations of D3-brane worldvolume fields from $r \to \infty$ towards the spike, and then ``measure'' what comes back to $r \to \infty$. In CFT terms, in the effective theory valid below the W-boson mass scale, $SU(N-1)$ $\N=4$ SYM plus $U(1)$ $\N=4$ SYM, we will scatter waves of $U(1)$ $\N=4$ SYM degrees of freedom off the screened Wilson line. We will indeed find phase shifts and resonant scattering, producing peaks in cross sections. The positions and widths of these peaks will be determined by the real and imaginary parts of the QNM frequencies, respectively. Moreover, these peaks will have an asymmetric shape characteristic of Fano resonances~\cite{RevModPhys.82.2257}. Such Fano resonances clearly arise from the mechanism of refs.~\cite{Erdmenger:2013dpa,Erdmenger:2016vud,Erdmenger:2016jjg}, namely the breaking of $(0+1)$-dimensional conformal symmetry at the Wilson line.

We focus on the fluctuations $\rho_0$, with equation of motion in eq.~\eqref{eq:rho0eom}, and $B_l$, with equation of motion in eq.~\eqref{eq:abetaeom}. These equations take the generic form in eq.\eqref{eq:y},
\begin{equation}
\label{eq:y2}
    \partial_r^2 y(r) + \g_1(r) \partial_r y(r) + \le[ \w^2 \g_2(r) - \frac{l(l+1)}{r^2} \ri]  y(r) = 0.
\end{equation}
The asymptotics near the impurity, $r \to 0$, and asymptotically far away, $r \to \infty$, appear in eqs.~\eqref{eq:eomsmallr} and~\eqref{eq:eomlarger} for $\rho_0$ and eqs.~\eqref{eq:abetasmallr} and~\eqref{eq:abetalarger} for $B_l$, with the generic form
\beq
y(r)=\begin{cases}  \frac{c^y_{-l-1}}{r^{l+1}} \, \le[1 + \mathcal{O}\left(r\right)\ri] + d^y_l \, r^l \le[1  + \mathcal{O}\left(r\right) \ri], & r \to 0,\\ & \\
    f_l^y(r)\,e^{-i (\w r - l \pi/2)}+ g_l^y(r) \,e^{i (\w r - l \pi/2)}, & r \to \infty,
\end{cases}
\eeq
where $c^y_{-l-1}$, $d^y_l$, $f_l^y(r)$, and $g_l^y(r)$ are complex-valued. To shoot in a wave that scatters off the spike, and then measure the resulting phase shift of the wave that comes out, we will impose $g_l^y(r) \to -f_l^y(r) e^{2i \delta_l}$ as $r \to \infty$, where the phase $\delta_l$ in general depends on $l$, $Q$, and $\omega$. In contrast, for the QNMs we imposed a purely out-going boundary condition, $f_l^y(r)=0$ with non-zero $g^y_l(r)$, as mentioned at the end of sec.~\ref{sec:fluc}. Of course, for both the QNMs and the scattering solutions we impose normalisablity in the $r \to 0$ region, meaning $c^y_{-l-1}=0$.

Our objective is to determine the dependence of the phase shifts $\delta_l$ on $l$, $Q$, and $\omega$. To do so we use the variable phase method~\cite{calogero1967variable,1980JCoPh..38..327C}, as follows. We define two new functions, $a(r)$ and $\tilde{\delta}_l(r)$, via the ansatz
\begin{subequations}
\label{eq:variable_phase_ansatz}
\begin{align}
\label{eq:yansatz1}
    y(r) &= \frac{a(r)}{r} \left(
        e^{-i (\omega r - l \pi/2)} - e^{2 i \tilde{\delta}_l(r)} e^{i (\omega r - l \pi/2)}
    \right),
    \\
    \label{eq:yansatz2}
    \partial_r y(r) &= - i \omega \frac{a(r)}{r} \left(
        e^{-i (\omega r - l \pi/2)} + e^{2 i \tilde{\delta}_l(r)} e^{i (\omega r - l \pi/2)}
    \right).
\end{align}
\end{subequations}
In the asymptotic region $r \to \infty$ these new functions behave as $a(r) \to r \, f_l^y(r)$ and $\tilde{\delta}_l(r) \to \delta_l$, so to compute the phase shifts $\delta_l$ we need to solve for $\tilde{\delta}_l(r)$ and extract $\lim_{r \to \infty} \tilde{\delta}_l(r)$. We can obtain an equation for $\tilde{\delta}_l(r)$ as follows. Requiring the expression for $\partial_r y(r)$ in eq.~\eqref{eq:yansatz2} to be the derivative of the expression for $y(r)$ in eq.~\eqref{eq:yansatz1} allows us to solve for $\partial_r a(r)$ in terms of $a(r)$, $\tilde{\delta}_l(r)$, and $\partial_r \tilde{\delta}_l(r)$,
\beq
\label{eq:a_prime_solution}
    \partial_r a(r) = a(r)\le[\frac{1}{r} - i \partial_r\tilde{\delta}_l(r) + \cot \le( \frac{l \pi}{2} - \w r - \tilde{\delta}_l(r)\ri) \partial_r\tilde{\delta}(r) \ri].
\eeq
We then plug the ansatz in eq.~\eqref{eq:variable_phase_ansatz} into the equation of motion eq.~\eqref{eq:y2} and use eq.~\eqref{eq:a_prime_solution} to eliminate $\partial_ra(r)$. The resulting equation has an overall factor of $a(r)$, however a non-trivial solution will have $a(r)\neq0$, leaving us with an equation for $\tilde{\delta}_l(r)$,
\begin{multline}
\label{eq:variable_phase_eom}
    \omega \, \partial_r\tilde{\delta}(r) + \frac{1}{2} \w \gamma_1(r) \sin \le(
        l \pi - 2 \w r - 2 \tilde \delta(r)
    \ri)
    \\
    + \frac{l(l+1) + \w^2 r^2 (1 - \gamma_2(r))}{r^2} \sin^2 \le( \frac{l \pi}{2} - \w r - \tilde\delta(r) \ri) = 0.
\end{multline}
To compute the $\delta_l$ we thus need to solve eq.~\eqref{eq:variable_phase_eom}. We do so via numerical shooting with parameter $\omega$, which unlike the QNM case we now restrict to real values. Specifically, we dial through $\omega$ values, for each value solving eq.~\eqref{eq:variable_phase_eom} with the boundary condition $c^y_{-l-1}=0$ to guarantee normalisability at $r \to 0$, and then we extract $\lim_{r \to \infty} \tilde{\delta}_l(r) = \delta_l$. We can translate $c^y_{-l-1}=0$ into a boundary condition on $\tilde{\delta}_l(r)$ by expanding the ansatz in eq.~\eqref{eq:variable_phase_ansatz} around $r=0$ and demanding that $a(r)$ and $\tilde{\delta}_l(r)$ are regular there, which gives
\beq
c^y_{-l-1} = e^{-i l \pi/2} \left(e^{i l \pi} - e^{2 i \tilde{\delta}_l(0)}\right) a(0).
\eeq
We thus obtain $c^y_{-l-1}=0$ by imposing $\tilde{\delta}_l(0)=\frac{\pi}{2} \,l$.

\begin{figure}
\begin{subfigure}{0.5\textwidth}
		\includegraphics[width=\textwidth]{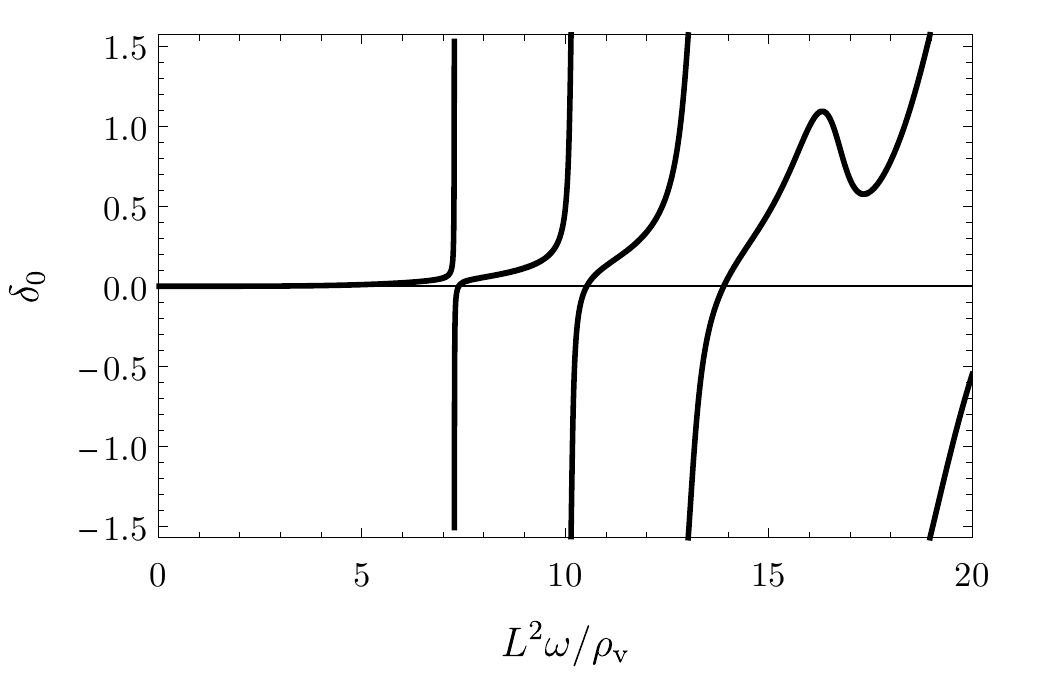}
		\caption{$Q = 0.01$, phase shift}
	\end{subfigure}
	\begin{subfigure}{0.5\textwidth}
		\includegraphics[width=\textwidth]{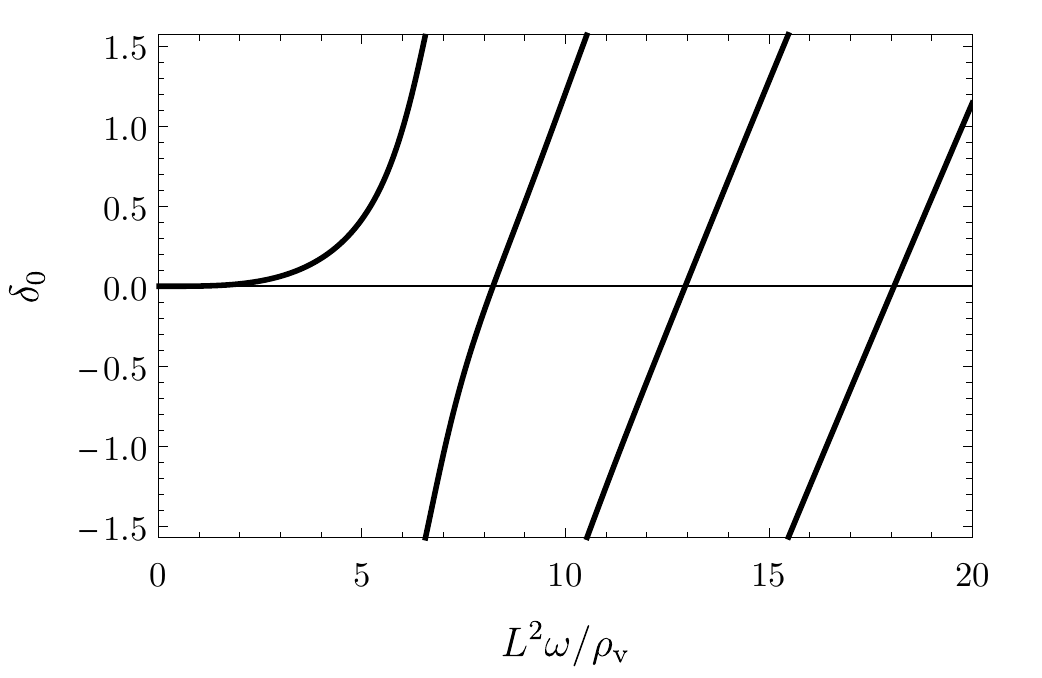}
		\caption{$Q = 0.1$, phase shift}
	\end{subfigure}
	\begin{subfigure}{0.5\textwidth}
		\includegraphics[width=\textwidth]{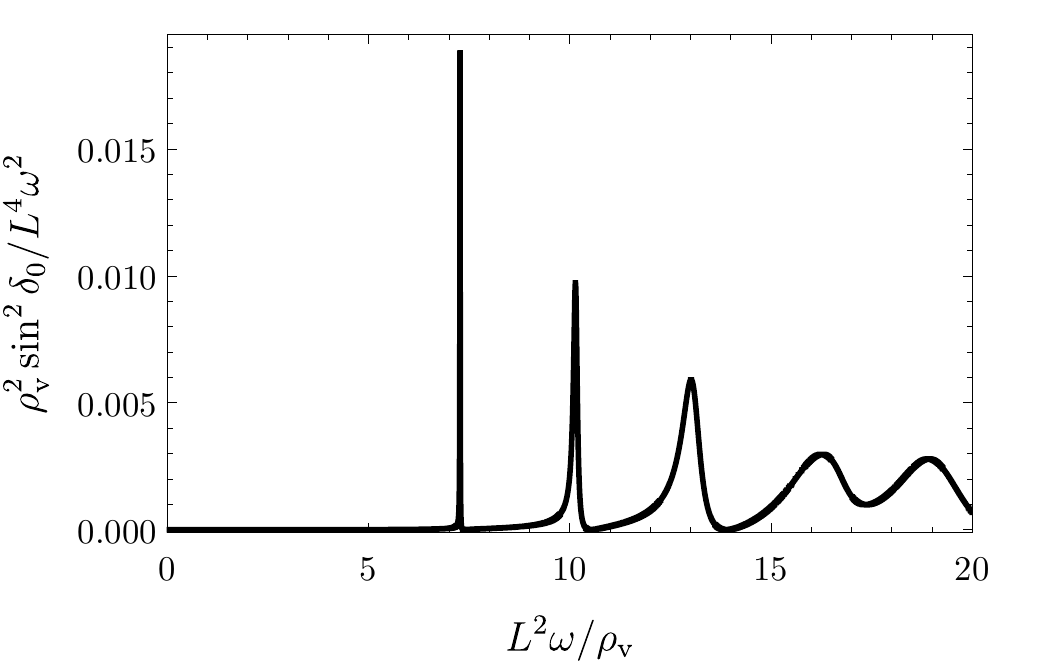}
		\caption{$Q = 0.01$, cross section}
	\end{subfigure}
	\begin{subfigure}{0.5\textwidth}
		\includegraphics[width=\textwidth]{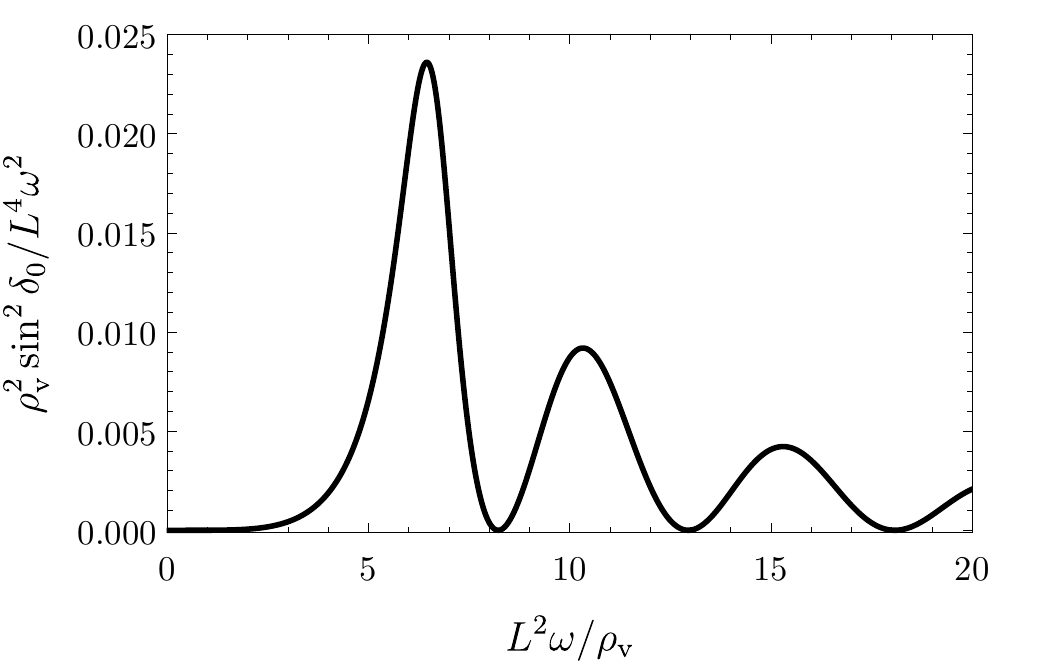}
		\caption{$Q = 0.1$, cross section}
	\end{subfigure}
	\begin{subfigure}{0.5\textwidth}
		\includegraphics[width=\textwidth]{s_wave_qnm_k0p01}
		\caption{$Q = 0.01$, QNMs}
	\end{subfigure}
	\begin{subfigure}{0.5\textwidth}
		\includegraphics[width=\textwidth]{s_wave_qnm_k0p1}
		\caption{$Q = 0.1$, QNMs}
	\end{subfigure}
	\caption{(a) and (b) The phase shift $\delta_0$ for the fluctuation $\rho_0$ as a function of real-valued frequency $\omega$ for $Q = 0.01$ and $0.1$. (c) and (d) The associated cross section $\sin^2\left(\delta_0\right)/\omega^2$, in units of $L^4/\rhov^2$, as a function of real-valued frequency $\omega$, for $Q = 0.01$ and $0.1$. (e) and (f) The complex $\omega$ plane, where black crosses denote QNMs of $\rho_0$, for $Q=0.01$ and $0.1$ (copied from fig.~\ref{fig:s_wave_complex_plane1}). In all plots $\omega$ is in units of $\rhov/L^2$.}
	\label{fig:s_wave_phase_shift}
\end{figure}

Given a scattering phase shift $\delta_l$ as a function of $\omega$, in analogy with quantum mechanics we can define a scattering cross section proportional to $\sin^2\left(\delta_l\right)/\omega^2$. For the fluctuation $\rho_0$, fig.~\ref{fig:s_wave_phase_shift} shows the phase shift $\delta_0$ and cross section as functions of (real-valued) $\omega$, and also, for comparison, the QNMs from fig.~\ref{fig:s_wave_complex_plane1}. The phase shift clearly changes rapidly at certain $\omega$, leading to resonances in the cross section whose positions and widths are determined by the real and imaginary parts of the QNM frequencies, respectively. In other words, if a QNM appears at some point $(\textrm{Re}\left(\omega\right),\textrm{Im}\left(\omega\right))$ in the complex $\omega$ plane, then the phase shift exhibits rapid variation near that $\textrm{Re}\left(\omega\right)$, and the corresponding cross section exhibits a peak there, of width $\propto 2|\textrm{Im}\left(\omega\right)|$, which is obvious when comparing the three figures in either column of fig.~\ref{fig:s_wave_phase_shift}. Such behaviour is familiar from quantum mechanics, where rapidly-changing phase shifts and resonances in cross sections indicate quasi-bound states of the Schr\"odinger potential.

On general grounds we expect the resonances in the cross section to have a Fano line-shape~\cite{RevModPhys.82.2257}. A Fano resonance arises whenever a standard Lorentzian resonance, characterised by a position and width, is coupled to a continuum of modes (in energy). If we scatter modes from the continuum off the resonance, they have two options: either interact with the resonance (resonant scattering) or not (non-resonant scattering). The Fano lineshape is thus characterised not only by a position, height, and width, but also by the Fano parameter, $q$, where $q^2$ is proportional to the ratio of probabilities of resonant to non-resonant scattering. A finite value of $q$ changes the resonance's line-shape from symmetric (Lorentzian) to asymmetric (Fano). Specifically, the cross section of a Fano resonance is
\begin{equation}
	\sigma_{\textrm{Fano}} = \sigma_0 \, \frac{(q \Gamma/2 + \omega - \omega_0)^2}{(\Gamma/2)^2 + (\omega - \omega_0)^2},
	\label{eq:fano}
\end{equation}
with normalisation $\sigma_0$, position $\omega_0$, and width $\Gamma$. The Fano line-shape reduces to a Lorentzian when $q \to \infty$, which means infinite coupling between resonance and continuum.

\begin{figure}
	\begin{center}
	\includegraphics[width=0.9\textwidth]{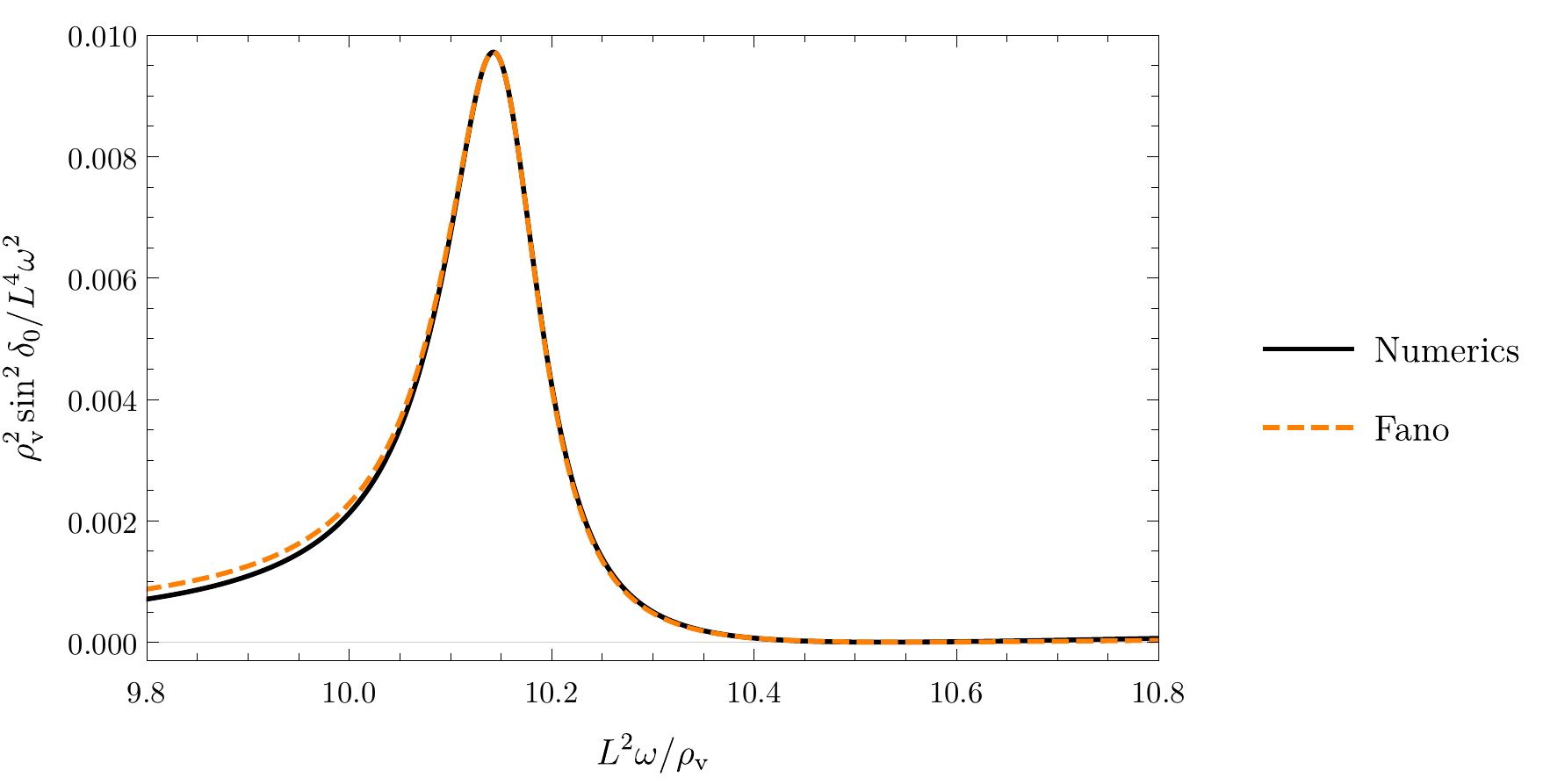}
\end{center}
	\caption{Close-up of the second resonance from the left in fig.~\ref{fig:s_wave_phase_shift} (c), for fluctuations of $\rho_0$ with $Q = 0.01$. The black solid line is our numerical result, while the orange dashed line is the Fano line shape from eq.~\eqref{eq:fano} with position $\omega_0$ and width $\Gamma$ determined from the nearest QNM frequency to be $\omega_0 - i \Gamma/2 \approx (10.2 - 0.0573 i) 
	\rhov/L^2$. Matching the position and height of the peak then gives normalisation $\s_0\approx 1.96 \times 10^{-4} L^4/\rhov^2$ and Fano parameter $q \approx -6.96$.}
		\label{fig:closeup}
\end{figure}

A key observation of refs.~\cite{Erdmenger:2013dpa,Erdmenger:2016vud,Erdmenger:2016jjg} was that systems in which $(0+1)$-dimensional conformal symmetry is broken, for example by an RG flow induced by a relevant operator or by an operator VEV, will generically give rise to Fano resonances. The reason is simple. Conformal symmetry implies a continuum of modes, for example any spectral function must be simply a power of $\omega$ determined by dimensional analysis, and hence must be a featureless continuum. Producing a resonance with some $\omega_0$ and $\Gamma$ obviously requires breaking conformal symmetry. Crucially, in $(0+1)$ dimensions the continuum cannot avoid the resonance, unlike higher dimensions where modes from the continuum can avoid the resonance either in real space (i.e. large impact parameter) or in momentum space. In $(0+1)$ dimensions the continuum has no place to escape the resonance, and the result is therefore Fano line-shapes.

\begin{figure}
\begin{subfigure}{0.5\textwidth}
    \includegraphics[width=\textwidth]{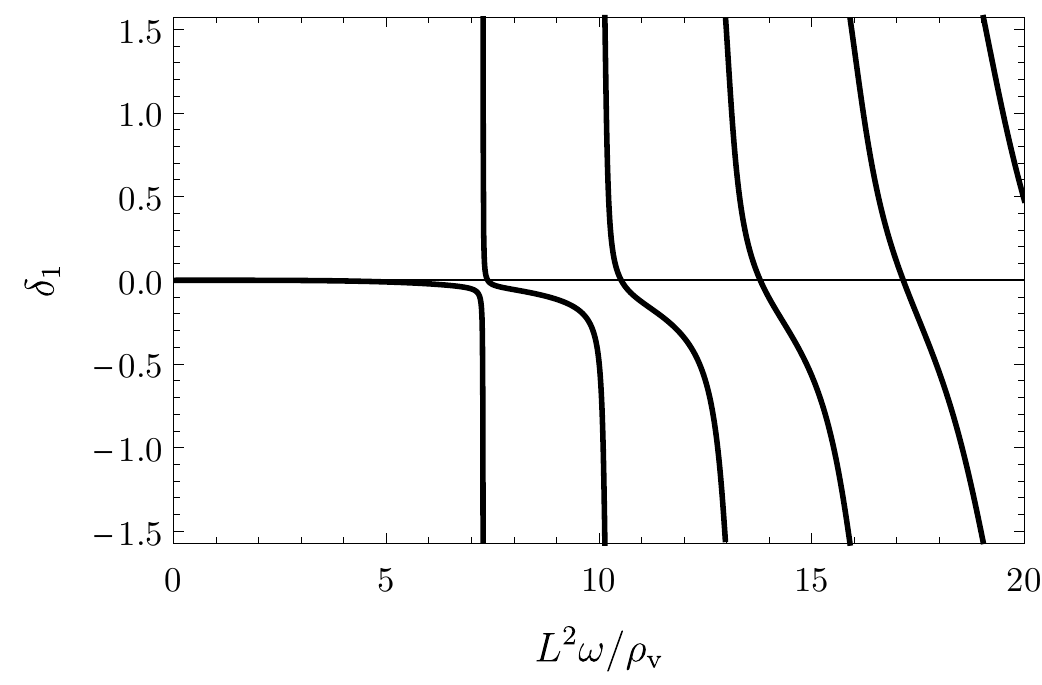}
    \caption{$Q = 0.01$, phase shift}
\end{subfigure}
\begin{subfigure}{0.5\textwidth}
    \includegraphics[width=\textwidth]{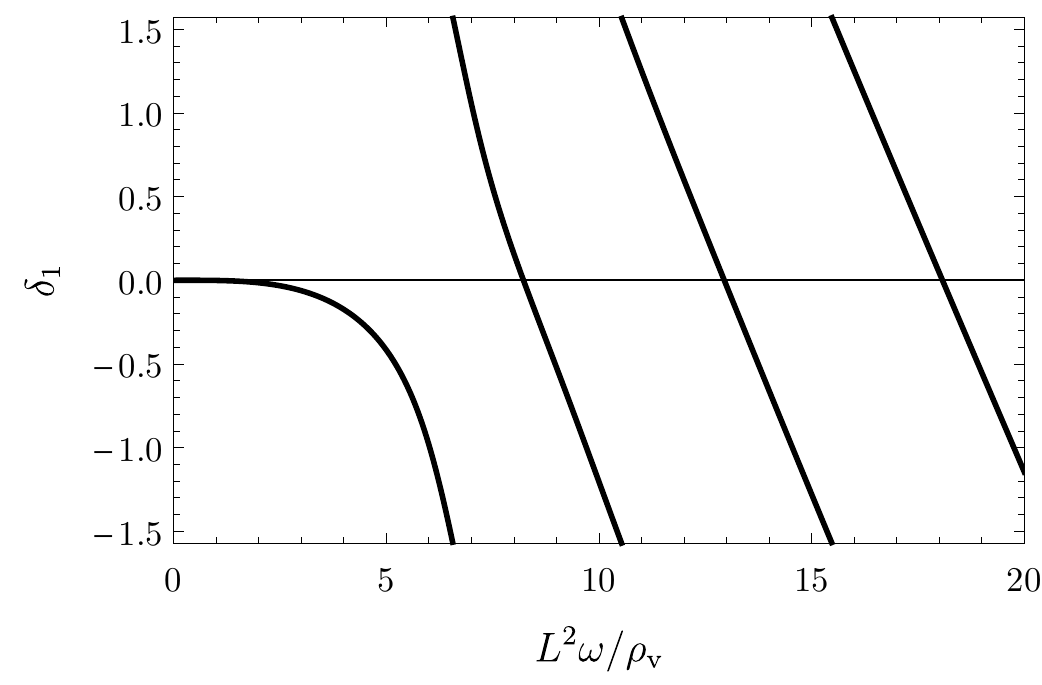}
    \caption{$Q = 0.1$, phase shift}
\end{subfigure}
\begin{subfigure}{0.5\textwidth}
    \includegraphics[width=\textwidth]{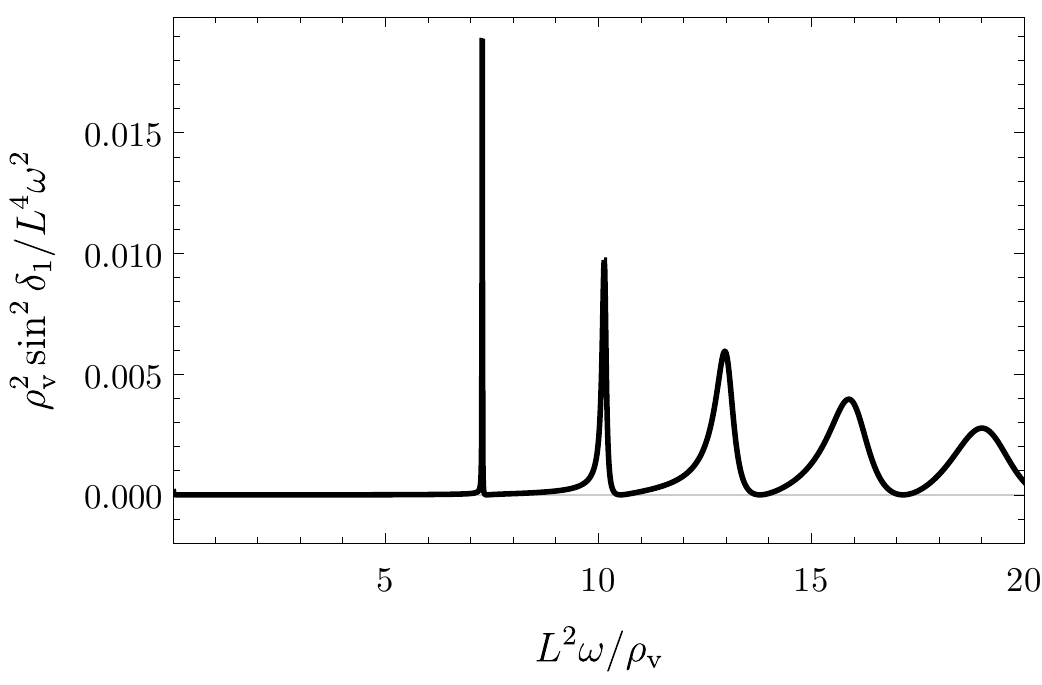}
    \caption{$Q = 0.01$, cross section}
\end{subfigure}
\begin{subfigure}{0.5\textwidth}
    \includegraphics[width=\textwidth]{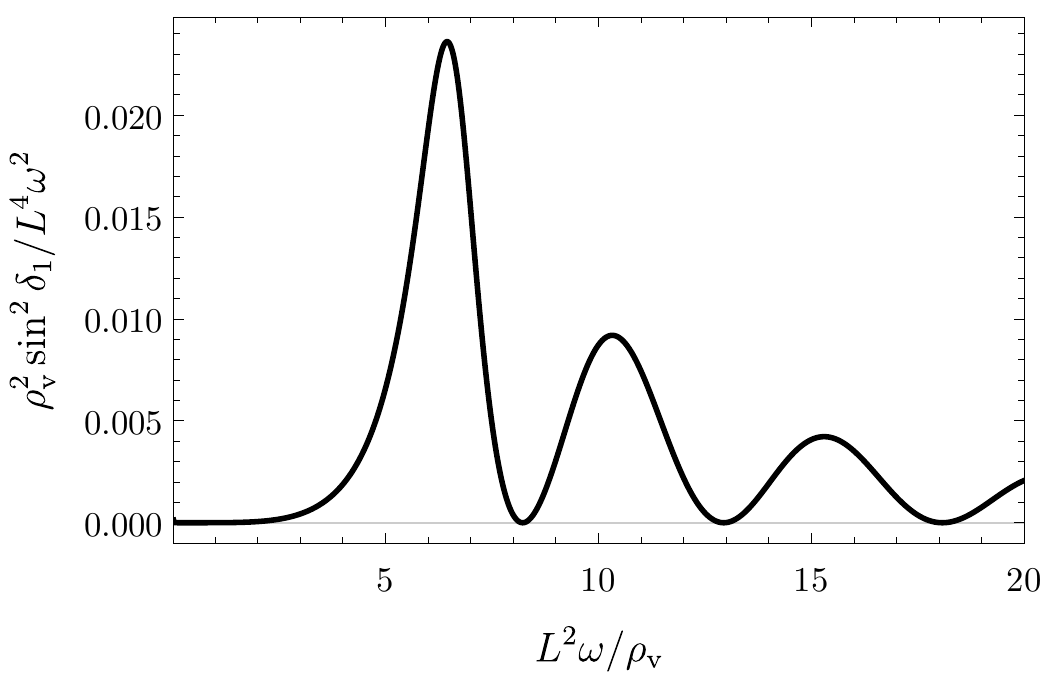}
    \caption{$Q = 0.1$, cross section}
\end{subfigure}
\begin{subfigure}{0.5\textwidth}
    \includegraphics[width=\textwidth]{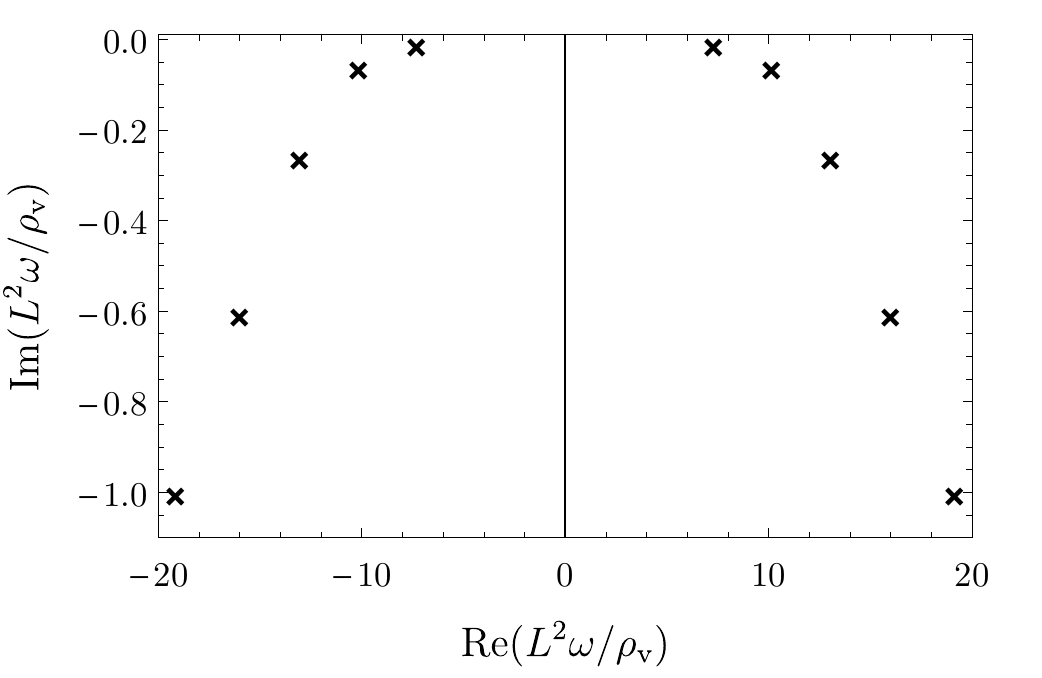}
    \caption{$Q = 0.01$, QNMs}
\end{subfigure}
\begin{subfigure}{0.5\textwidth}
    \includegraphics[width=\textwidth]{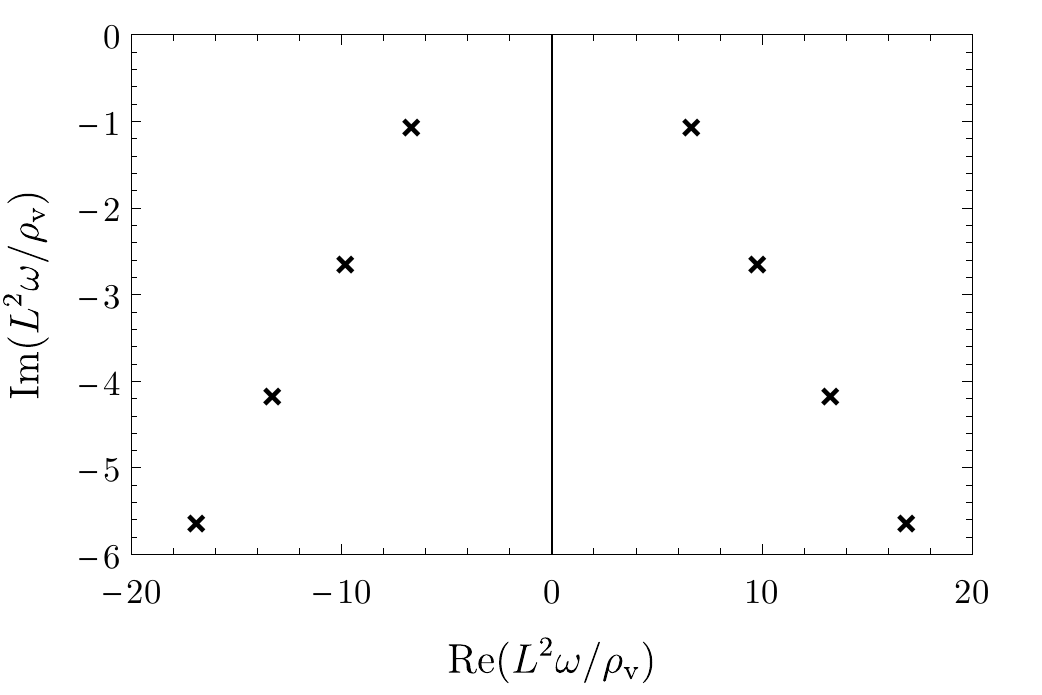}
    \caption{$Q = 0.1$, QNMs}
\end{subfigure}
\caption{(a) and (b) The phase shift $\delta_1$ for the fluctuation $B_l$ with $l=1$ as a function of real-valued frequency $\omega$ for $Q = 0.01$ and $0.1$. (c) and (d) The associated cross section $\sin^2\left(\delta_1\right)/\omega^2$, in units of $L^4 / \rhov^2$, as a function of real-valued frequency $\omega$, for $Q = 0.01$ and $0.1$. (e) and (f) The complex $\omega$ plane, where black crosses denote QNMs of $B_l$ with $l=1$, for $Q=0.01$ and $0.1$ (copied from fig.~\ref{fig:qnm_beta_multiple_l}). In all plots $\omega$ is in units of $\rhov/L^2$.}
\label{fig:p_wave_beta_sample_k}
\end{figure}

As mentioned in sec.~\ref{sec:spike}, for the cone D3-brane the $AdS_2$ factor in the worldvolume geometry indicates $(0+1)$-dimensional conformal symmetry, which is then broken in the spike solution, hence by the arguments above we expect Fano line-shapes. Numerically we indeed find that the cross section resonances in fig.~\ref{fig:s_wave_phase_shift} are of Fano form. For example, fig.~\ref{fig:closeup} shows a fit of the Fano form in eq.~\eqref{eq:fano} to the second peak from the left in fig.~\ref{fig:s_wave_phase_shift} (c), with excellent agreement, where $\omega_0$ and $\Gamma$ are determined by the real and imaginary parts of the associated QNM frequency, respectively, and numerically we find $q \approx -6.96$.

Fig.~\ref{fig:p_wave_beta_sample_k} shows our numerical results for the phase shift, cross section, and QNMs for p-wave scattering off the impurity, namely for the fluctuation $B_l$ with $l=1$. These results are qualitatively similar to the s-wave case in fig.~\ref{fig:s_wave_phase_shift}, namely the phase shifts vary rapidly at the real parts of QNM frequencies, and the cross sections have Fano resonances at the same frequencies, with widths determined by the imaginary parts of QNM frequencies. Indeed, we expect these features to be generic to all fluctuations of D3-brane worldvolume fields.

\section{Summary and Outlook}
\label{sec:outlook}

In this paper we re-visited the well-known ``spike'' solutions for probe D3-branes in $AdS_5 \times S^5$, holographically dual to a symmetric-representation Wilson line on the Coulomb branch of $\N=4$ SYM~\cite{Gauntlett:1999xz,deMelloKoch:1999ui,Ghoroku:1999bc,Drukker:2005kx,Fiol:2008gt,Schwarz:2014rxa,Schwarz:2014zsa}. We have presented compelling evidence that these solutions describe a Wilson line \textit{screened} by the adjoint scalar VEV, in a fashion similar to impurity screening in a LFL. Intuitively, we imagine a spherically-symmetric cloud of the adjoint scalar VEV that behaves as a collection of color dipoles polarized by the Wilson line ``impurity,'' and reducing its effective ``charge'' to zero at spatial infinity. In particular, by solving for linearised fluctuations of bosonic D3-brane worldvolume fields, we showed that the Wilson line impurity supports quasi-bound states, scattering phase shifts, and Fano resonances in scattering cross sections, just like an impurity screened by a LFL. Furthermore, we claim that the mechanisms for these phenomena will be generic to screened impurities in holography, as they arise simply from the fact that bulk modes can scatter off the localised spike of D3-brane and then escape to infinity, producing QNMs even without a black hole horizon.

Our results suggest many avenues for further research. For example, we considered only a subset of the bosonic fluctuations of the spike D3-brane worldvolume fields. What about the other bosonic fluctuations, or the fermionic fluctuations?

In a LFL, a key feature of screened impurity physics is the Friedel sum rule~\cite{doi:10.1080/14786440208561086}. The change in the electronic spectral function due to the impurity measures the average charge bound to the impurity, or intuitively the total number of bound electrons, which in turn determines $\sum_l (2l+1) d\delta_l/d\omega$, with phase shifts $\delta_l$. Integrating up to the Fermi level then gives the Friedel sum rule: the number of bound electrons is proportional to $\sum_l \delta_l$, which is typically approximated at low energies as simply $\delta_0$. The Friedel sum rule comes from standard LFL Ward identities and hence holds for any strength of coupling between the impurity and LFL electrons, or between the impurity electrons themselves (Coulomb repulsion)~\cite{PhysRev.121.1090,PhysRev.150.516}. Intuitively, Friedel's sum rule is a ``node counting theorem.'' If $\delta_0>0$ then the LFL quasi-particles are drawn inwards, towards the impurity, and every time $\delta_0$ passes through $\pi$ the LFL quasi-particle wave function at infinity acquires a new node, signalling that another unit of charge has been ``lost,'' i.e. another electron has become bound to the impurity~\cite{Coleman2015}. What happens to the Friedel sum rule when the LFL is replaced by strongly-interacting degrees of freedom, with no quasi-particle description, is an open question.

Presumably, for the worldvolume fields of the D3-brane spike, the change in spectral functions should similarly determine $\sum_l (2l+1) d\delta_l/d\omega$. We computed both the spectral functions of our bosonic excitations and their $d\delta_l/d\omega$, but found no obvious relation between the two. We can venture an explanation for why, namely a key difference between a LFL and our system. In a LFL, both the impurity and the electrons are charged under the $U(1)$ of electromagnetism and/or the $SU(2)$ of spin. In our system, on the other hand, the spike is charged under the D3-brane's $U(1)$ worldvolume gauge field, but the worldvolume fields we scattered off the spike are not: in the Abelian DBI action the worldvolume fields couple non-linearly to the $U(1)$ gauge field, but none of them are charged, i.e. none of them have a covariant derivative with respect to the $U(1)$ gauge field. Our scattering is therefore more similar to scattering light off of a charged impurity than scattering electrons, albeit in strongly-interacting $\N=4$ SYM rather than Maxwell theory. In any case, while quasi-bound states, phase shifts, and cross section resonances are generic to impurities, the nature of the Friedel sum rule clearly depends on what is being scattered.

Simple generalisations of the D3-brane spike can describe a variety of other single-impurity systems. For example, worldvolume $SL(2,\mathbb{R})$ transformations can convert the electrically-charged spike into a magnetically-charged spike, or more generally a dyonic spike. In CFT terms, $SL(2,\mathbb{R})$ transformations can convert the Wilson line into an 't Hooft line, or more generally a mixed Wilson-t' Hooft line. Furthermore, sending $Q \to -Q$ effectively ``flips'' the spike, so that now instead of extending up to the $AdS_5$ boundary it extends down to the Poincar\'e horizon. Such solutions describe W-bosons, or after $SL(2,\mathbb{R})$ transformations, magnetic monopoles, and more generally dyonic excitations of the Coulomb branch (see for example refs.~\cite{Schwarz:2013wra,Schwarz:2014rxa,Schwarz:2014zsa}). All of these can be characterised by their spectra of QNMs, and associated phase shifts and cross sections. The magnetically charged solutions break parity symmetry, which allows for new couplings among worldvolume fluctuations, so we expect that their QNMs will indeed be different from those of the electric spike.

In fact, another quantity that may be crucial for characterising such impurities is entanglement entropy, which can measure impurity entropy, i.e. the impurity's ground state degeneracy~\cite{Calabrese:2004eu}. For a spherical entangling region centered on the impurity, the change in entanglement entropy due to a Wilson line, both screened and not, and due to a W-boson, has been computed in refs.~\cite{Lewkowycz:2013laa,Kumar:2017vjv,Kobayashi:2018lil}. Such entanglement entropy measures the amount of correlation between the impurity and bulk degrees of freedom. For example, the spike describes an order $N$ entanglement entropy between the Wilson line and adjoint scalar VEV, while the separate cone and flat D3-branes have zero entanglement entropy at order $N$.

However, perhaps the most tantalising generalisation involves the multi-centre solutions mentioned at the end of sec.~\ref{sec:spike}. As mentioned there, these can describe \textit{lattices} of Wilson lines, or via the generalisations mentioned above, 't Hooft lines, mixed Wilson-'t Hooft lines, W-bosons, monopoles, etc. Sticking to Wilson lines for clarity, as we mentioned in sec.~\ref{sec:intro} the adjoint scalar VEV can screen these or not, at no cost in energy, so in fact we can construct a lattice with whatever mix we like of screened and un-screened Wilson lines. We choose the sites of the lattice by hand, so we can construct lattices in one, two, or three-dimensional sub-spaces of the $(3+1)$-dimensional $\N=4$ SYM, with whatever shape we like (square, hexagonal, etc.). In the language of AdS/CFT, such Wilson line lattices, and their $SL(2,\mathbb{R})$ generalisations, are ``non-normalisable,'' however we can also construct lattices of normalisable objects, namely the W-bosons and their $SL(2,\mathbb{R})$ cousins.

Crucially, all of these are exact 1/2-BPS solutions at $T=0$, which has advantages and disadvantages. Among the advantages are that we have a vast class of exact solutions in which translational symmetry is broken either explicitly, for example by Wilson lines, or spontaneously, for example by W-bosons. We can therefore calculate the spectrum of fluctuations of worldvolume fields, which via a holographic version of Bloch's theorem should give rise to a band structure. A key question then is whether that band structure exhibits gapless modes, including in particular phonons or topologically-protected gapless (edge) modes. In other words, can we use these types of solutions to construct a holographic topological insulator? To be clear, by ``insulator'' we imagine the $U(1)$ $\N=4$ SYM sector as ``electromagnetism,'' so that the only charged excitations are in the W-boson multiplet, which is gapped, hence the ground state is insulating. We can break discrete symmetries, such as time reversal, by using 't Hooft lines instead of Wilson lines, for example.

In many experimental systems, changing the concentration of impurities coupled to a LFL results in (quantum) phase transitions. For example, in a Kondo lattice of magnetic impurities coupled to a LFL, the competition between Kondo and RKKY interactions gives rise to a quantum critical point~\cite{Coleman2015}. The associated finite-$T$ quantum critical phase is the so-called strange metal state, which has an electrical resistivity $\propto T$, unlike a LFL's $T^2$.

Our holographic D3-brane lattices cannot exhibit any such quantum phase transition: they are BPS and hence the impurities obey a no-force condition. No competition between interactions is present to produce a change in symmetry. Additionally, these BPS solutions are only known exactly at $T=0$, and indeed a non-zero $T$ will generically break SUSY, making exact solutions much more difficult to obtain. In fact, a $T>0$ version of the cone D3-brane solution may not exist~\cite{Hartnoll:2006hr}. As a result, the existing exact BPS solutions cannot be used to study the finite-$T$ behavior of observables such as electrical resistivity.

Nevertheless, despite these disadvantages, these classes of probe D3-brane solutions clearly offer a vast array of important and worthwhile opportunities, which we intend to pursue, using the results of this paper as a foundation.

\acknowledgments

We thank C.~Bachas, J.~Erdmenger, C.~Hoyos, and S.~P.~Kumar for useful discussions and correspondence. We especially thank M.~J\"arvinen for reading and commenting on the manuscript. We acknowledge support from STFC through Consolidated Grant ST/P000711/1. A.~O'B. is a Royal Society University Research Fellow. The work of R.~R. was supported in part by the D-ITP consortium, a program of the Netherlands Organisation for Scientific Research (NWO) that is funded by the Dutch Ministry of Education, Culture and Science (OCW). A.~O'B. and R.~R. thank the Institut Henri Poincar\'e for hospitality while this work was in progress.

\bibliographystyle{JHEP}
\bibliography{d3spike}

\end{document}